\documentclass[showpacs,aps,amsmath,letterpaper,nofootinbib,reprint,showkeys,superscriptaddress,pra]{revtex4-2}
\usepackage{dcolumn}    
\usepackage{multirow}
\usepackage{bm}         
\usepackage{ifpdf}
\usepackage{amssymb,lineno,amsfonts}
\usepackage{graphicx}   
\usepackage{amsmath}
\usepackage{bbm}
\usepackage{comment}
\usepackage{mathrsfs}
\usepackage{mathtools}
\usepackage{booktabs} 
\usepackage{colortbl} 
\usepackage{upgreek}
\usepackage[table]{xcolor}
\usepackage{physics}
\usepackage{amsthm}
\usepackage{epstopdf}
\usepackage{setspace}
\usepackage{dashbox}
\usepackage[toc,page]{appendix}
\usepackage{hyperref}
\usepackage{cleveref}
\usepackage{bbold}
\usepackage[all]{xy}
\usepackage{soul}
\usepackage{tikz}
\usepackage{natbib}
\usepackage{color} 
\usepackage{subfigure}
\usepackage{bbold}
\usepackage{tikz}

\usepackage{amsthm}
\usepackage{academicons}
\usepackage[english]{babel}

\definecolor{med-blue}{RGB}{25,25,112} 
\hypersetup{colorlinks, linkcolor={blue},citecolor={blue}, urlcolor={blue}}

\newcommand{\proj}[1]{\vert{#1}\rangle \langle{#1}\vert}
\newcommand{\Ket}[1]{\vert{#1}\rangle}
\newcommand{\Bra}[1]{\langle{#1}\vert}

\usepackage{orcidlink}

\newcommand{\expv}[2]{\left\langle{#1}\right\rangle_{#2}}
\newcommand{\dmel}[2]{\left\langle{#1}\left\vert{#2}\right\vert{#1}\right\rangle}
\begin{document}
\title{Partial Quantum Shadow Tomography for Structured Operators \\ and its Experimental Demonstration using NMR
} 

\author{Aniket Sengupta\orcidlink{0009-0007-2522-6718}}
\email{aniket.sengupta@students.iiserpune.ac.in}
\affiliation{Department of Physics, Indian Institute of Science Education and Research, Pune 411008, India}
\affiliation{NMR Research Center, Indian Institute of Science Education and Research, Pune 411008, India}

\author{Arijit Chatterjee\orcidlink{0009-0000-8191-1882}}
\email{arijitchattopadhyay01@gmail.com}
\affiliation{Department of Physics, Indian Institute of Science Education and Research, Pune 411008, India}
\affiliation{NMR Research Center, Indian Institute of Science Education and Research, Pune 411008, India}

\author{G. J. Sreejith\orcidlink{0000-0002-2068-1670
}}
\email{sreejith@acads.iiserpune.ac.in}
\affiliation{Department of Physics, Indian Institute of Science Education and Research, Pune 411008, India}

\author{T. S. Mahesh}
\email{mahesh.ts@iiserpune.ac.in}
\affiliation{Department of Physics, Indian Institute of Science Education and Research, Pune 411008, India}
\affiliation{NMR Research Center, Indian Institute of Science Education and Research, Pune 411008, India}
\def\thefootnote{$*,\dagger$}\footnotetext{These authors contributed equally to this work}\def\thefootnote{\arabic{footnote}}

\begin{abstract}
{Quantum shadow tomography based on the classical shadow representation provides an efficient way to estimate properties of an unknown quantum state without performing a full quantum state tomography. 
In scenarios where estimating the expectation values for only certain classes of observables is required, obtaining information about the entire density matrix is unnecessary. 
We propose a partial quantum shadow tomography protocol that estimates a subset of density matrix elements relevant to the expectation values of structured observables. Specifically, we identify specific subsets of the tomographically complete set ${\mathrm{Cl}}(2)^{\otimes n}$ and a simple pseudo-inverse of the associated channel, which can be used to estimate all elements of the density matrix with the same active order.
By restricting the protocol to smaller subsets of single-qubit Pauli measurements, it becomes experimentally more efficient.
We demonstrate the advantage over unitary designs, such as the Clifford, full Pauli basis, and methods utilizing mutually unbiased bases, by analytically deriving error bounds and numerically evaluating the protocol on structured operators.
We experimentally demonstrate the partial shadow estimation scheme for a wide class of two-qubit states (pure, entangled, and mixed) in the nuclear magnetic resonance (NMR) platform. 
The full density matrix, reconstructed experimentally by combining different partial estimators, achieves fidelities around 99\%.
}
\end{abstract}

\maketitle	

\section{Introduction} \label{sec:intro}
Quantum Shadow Tomography (QST)~\cite{Aaranson, Huang2020} is an efficient way to estimate a wide range of properties of an unknown quantum state via the data collected from projective measurements on the state.
Classical shadows have found applications in quantum simulation tasks such as probing quantum scrambling \cite{scrambling, scrambleCS}, in quantum machine learning tasks~\cite{Cho2024, Jerbi2024}, and vast usage in randomized measurement protocols for fidelity estimation, characterization of topological order~\cite{Elben2023}, energy estimation~\cite{hadfield2021adaptivepaulishadowsenergy}, entanglement detection~\cite{elben2020,Neven2021} and many more.
QST was motivated by the seminal work of Aaronson~\cite{Aaranson}, which established theoretical bounds for sampling complexity using Haar-random unitaries~\cite{AAMele}. Subsequent advances introduced the classical shadow framework~\cite{Huang2020}, replacing Haar-random unitaries with simpler unitary designs, such as the Clifford group~\cite{clifford3design, Multiqubitcliffordgroups} allowing for more efficient experimental implementations \cite{experimentQST, ExperimentalQuantumproperties}, particularly on near-term quantum devices. 
Apart from these two, other unitary ensembles which have been explored include fermionic Gaussian unitaries~\cite{Zhao_2021}, Pauli-invariant unitary ensembles~\cite{Bu2024}, and unitary ensembles corresponding to the time evolution of a random Hamiltonian~\cite{Hamil_Shadow}. Unitary ensembles defined through locally scrambled quantum dynamics~\cite{local_scramble_dynamics} have been shown to achieve a lower tomography complexity compared to Clifford-based methods. 
Recent developments in entangled bases measurements have shown a quadratic improvement in sampling complexity~\cite{Ippoliti2024classicalshadows} to learn Pauli expectation values. Classical shadows using mutually unbiased bases (MUBs)~\cite{Wang2024} provide a framework for shadow tomography for $n$-qubit systems by measuring along \(2^n+1\) MUBs, ensuring a robust sampling complexity. The construction of MUB circuits, as detailed in~\cite{yu2024efficientquantumcircuitconstruction}, employs a \(-cz-S-H-\) structure, enabling an efficient decomposition of each MUB circuit using \(O(n^2)\) gates within \(O(n^3)\) time. These advancements significantly reduce the required unitary samples and strive toward a systematic framework for efficient tomography.

Novel techniques in shadow tomography have been designed to mitigate the impact of experimental imperfections~\cite{NoiseShadowestimation, DaX}.
Neural networks have been used in combination with shadow tools for efficient quantum state reconstruction~\cite{NNQST}, which provides considerable advantages over direct shadow estimation. Its continued development focuses on optimizing protocols for scalability and noise resilience~\cite{Generalmeasurementframes, shadowinversion, Zhou2024efficientclassical, Guehne, Wu2024}.

Shadow tomography proceeds by measuring the quantum state (to be estimated) along a sufficiently rich set of directions in the full Hilbert space, with the directions parametrized by a set of unitaries that rotate the directions to $\hat{z}$. The richness of the unitary set guarantees that the protocol amounts to a quantum channel that is invertible. 
Smaller subsets of unitaries may not result in an invertible quantum channel, but they may admit pseudo-inverses that allow estimation of some parts of the original density matrix. Partial shadow tomography involves the identification of a subset of a tomographically complete set, an associated pseudo-inverse, and a prescription of a subspace in which the pseudo-inverse faithfully produces the matrix elements. Here we study such a partial quantum shadow tomography (PQST) protocol. We group the matrix elements of the density matrix with fixed active order. We identify subsets of ${\mathrm {Cl}}(2)^{\otimes n}$ unitaries, along with an associated pseudo-inverse map, which allow efficient estimation of matrix elements of a given active order.
This reduces unitary sampling complexity and improves estimation of expectation values for special cases of structured operators. We experimentally demonstrate PQST in an NMR system, where PQST, when combined with diagonal tomography of the ensemble system~\cite{DiagonalTomo}, can achieve exact estimation of density matrix elements, thereby offering significant advantages over shadow protocols in these contexts. 

This article is organized as follows. In Sec. \ref{sec:th}, we develop the theory of PQST, demonstrate 2- and 3-qubit cases and propose a generalization for $n$-qubit systems in Sec.~\ref{generalprotocol}. In Sec.~\ref{Sec:StrOPr}, we demonstrate PQST for structured operators. In Sec.~\ref{Sec:Sample}, we analyze the sample complexity of the PQST protocol and compare it with Clifford and Pauli measurement schemes. Sec. \ref{sec:Num} compares the performance of PQST protocol numerically with shadow tomography methods utilizing Clifford designs, Pauli measurements, and MUBs and highlights the advantages of PQST for the case of structured operators. In Sec. \ref{sec:exp}, we experimentally implement PQST using a 2-qubit NMR system. Finally, we discuss further prospects and conclude in Sec. \ref{sec:conclude}.

\begin{figure*}
\centering
\includegraphics[width=0.98\textwidth]{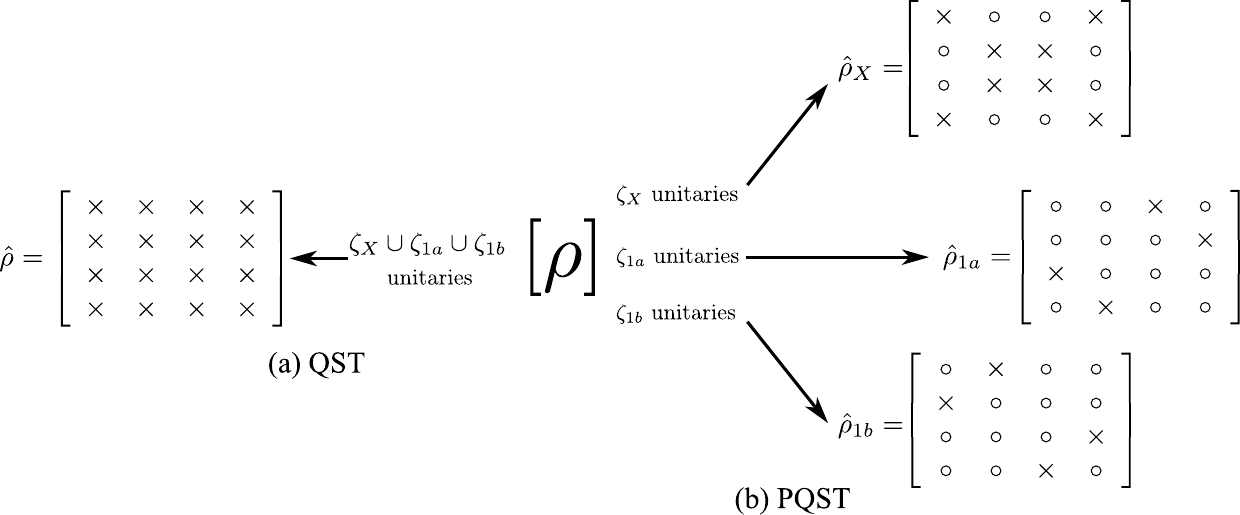} 
\caption{Illustrating (a) QST- full state characterization by sampling unitaries from a tomographic complete set of unitaries and (b) PQST-partial characterization by sampling unitaries from a subset of Clifford unitary design \(\mathrm{Cl(2)}^{\otimes n}\), on a 2-qubit system. In each partial shadow estimator $\widehat{\rho}_i$, the density matrix elements represented by \(\{\times\}\) are efficiently estimated in the PQST protocol, while the elements represented by \(\{\circ\}\) are discarded. However, by combining multiple partial shadow estimators $\{\widehat{\rho}_i\}$, the full density matrix is reconstructed, as described in Eq. \eqref{eq:rhorecover}.}
\label{fig:partial shadow}
\end{figure*}

\section{Theory} \label{sec:th}
\subsection{Quantum Shadow Tomography (QST)}
The general QST protocol is outlined below.
We perform a measurement procedure where we sample a unitary from the set $\zeta = \{U_i\}$ and apply it to the unknown state $\rho$, followed by a measurement in the computational basis $\ket{k}\in\{0,1\}^{ n}$. The set of unitaries needs to be sufficiently large that it is tomographically complete~\cite{Huang2020}- for any two distinct states, there should be at least one unitary $U\in\zeta$ and some computational basis state $b\in \proj{k}$ such that the two states have different expectation values for $U^\dagger\left| b\rangle \langle b\right |U$. Pauli basis measurements and the Clifford measurements are examples of such sets. The resulting collapsed state $\hat{\ket{k}}\hat{\bra{k}}$ is reverse rotated using the inverse unitary $U_i^\dagger$ and the resulting outcome $U_i^\dag \hat{\ket{k}}\hat{\bra{k}}  U_i $ is stored. We iterate this process 
\begin{equation}
\centering
    \rho \xrightarrow[\mathrm{rotate}]{U_i} \substack{\displaystyle U_i\rho U_i^\dag }\xrightarrow[\mathrm{Measure}]{} \hat{\ket{k}}\hat{\bra{k}}\xrightarrow[\substack{\text{Inverse} \\ \text{rotate}}]{U_i^\dag}  U_i^\dag  \hat{\ket{k}}\hat{\bra{k}} U_i 
\label{eq:shadow_0}
\end{equation}
for different choices of unitaries and measurement outcomes. Now, if we average $U_i^\dag  \hat{\ket{k}}\hat{\bra{k}} U_i $ over choices of unitary applications and measurement outcomes, we get a quantum channel map 
\begin{align}
    \rho\rightarrow \mathcal{M}(\rho)= {\mathbbm E}_{U,k}  [U^\dag  {\ket{k}}{\bra{k}} U]   \approx {{\mathbbm E}_{i,\hat{k}}}[U_i^\dag  \hat{\ket{k}}\hat{\bra{k}} U_i ]  
\label{eqn:Expectation}
\end{align}
where ${\mathbbm E}_{U,k}$ is the weighted average over $\zeta$ and the computational basis states $\vert k \rangle$ weighted by Born probabilities, which can be estimated using the empirical average ${\mathbbm E}_{i,\hat{k}}$ over both the sampled unitaries and the post measurement states $\hat{\ket{k}}$. The channel map has an inverse due to tomographic completeness of $\zeta$~\cite{Huang2020}. The shadow estimator of the density matrix of the original state is given by the action of this inverse on the reverse rotated measurement outcomes, averaged over the sampling size
\begin{equation}
\widehat{\rho} ={{\mathbbm E}_{i,\hat{k}}}[\mathcal{M}^{-1}({U_i^\dag  \hat{\ket{k}}\hat{\bra{k}} U_i })],
\end{equation}
$\widehat{\rho}$ is the shadow estimator, which in the limit of infinite shadows yields $\rho$~\cite{Huang2020}.
If we know $\mathcal{M}^{-1}$, we can retrieve the density matrix $\rho$ by taking an average over the classical shadows $\mathcal{M}^{-1}(U_i^\dag \hat{\ket{k}}\hat{\bra{k}} U_i)$. The quantum channel depends on the unitary ensemble. Sampling from the Haar measure over the full unitary group produces a depolarization channel given by 
 \begin{equation}
     \mathcal{D}_{1/2^n+1}(A)=\frac{A+\mathrm {Tr}(A)\mathbb{1}}{2^n+1}.
     \label{eqn:depolarization}
 \end{equation}
The inverse of the channel is given by 
 \begin{equation}
     \mathcal{D}_{1/2^n+1}^{-1}(A)=(2^n+1)A-\Tr(A)\mathbb{1}.
     \label{eq:chinv}
 \end{equation}
The same quantum channel is generated when unitaries are sampled uniformly from the Clifford group \(\mathrm{Cl}(2^n)\), consisting of \(
2^{n^2 + 2n} \prod\limits_{j=1}^{n} (4^j - 1)
\) unitaries for a $n$-qubit system.
Another tomographic complete set involves Pauli basis measurements, where the unitary operator set takes the form $U=\mathrm {Cl}(2)^{\otimes n}$. Effectively, the protocol amounts to making a sequence of random measurements in $x,y$ and $z$ directions picked independently on each site. The shadow estimator formed by the Pauli basis measurements is given by
\begin{align}
\widehat{\rho}={\mathbbm E}_{U\in\text{Cl}(2)^{\otimes n},\hat{k}}\left[\bigotimes\limits^n_{j=1}\mathcal{D}_{1/3}^{-1}(U_j^\dagger\proj{\hat{k}_j}U_j)\right],\nonumber\\
~~\text{where}~~\hat{k}_1,\dots,\hat{k}_n\in\{0,1\}
\label{eqn:pauliinv}
\end{align}  
In contrast to Clifford unitaries, Pauli basis measurements require only local control, with a significantly smaller set of unitaries (see Fig. \ref{fig:partial shadow} (a)).
\subsection{Estimation using subsets of tomographically complete unitaries}
QST provides full characterization of a quantum state and involves sampling unitaries from a tomographically complete set. 
This, in principle allows for the estimation of any observable and can be extended to quantities non-linear in the density matrix, such as the subsystem entropies.
In this work, we consider the task of estimating expectation values of observables which are Pauli strings \( \{\mathbb{1},X,Y,Z\}^{\otimes n} \), using shadow tomography with optimally chosen unitaries. We investigate whether sampling unitaries from subsets of a tomographically complete set still permits partial state reconstruction via the pseudo-inverse map defined as:
\begin{equation}  
    \mathcal{M}_p^{-1}(A) = pA - \mathrm{Tr}(A)\mathbb{1},  
    \label{eqn:Inverse}
\end{equation}  
where $p$ is the strength of the pseudo-inverse map. This map is not completely positive and trace-preserving. However, when \( p = 2^n + 1 \), it acts as an inverse depolarizing map that preserves the trace for states with \( \Tr(A) = 1 \), though it remains non-completely positive.  
It is called the pseudo-inverse because it serves as the inverse only for a subset of density matrix elements, enabling the selective estimation of those elements, as described in the following sections.

More precisely, we aim to determine suitable combinations of triplets \(\{p, \zeta', \mathcal{O}\}\), where \( p\) is the strength of the pseudo-inverse map \(\mathcal{M}_p^{-1}(\cdot) \) defined in Eq.~\eqref{eqn:Inverse}, \( \zeta'\) is a subset of the tomographically complete set \(\zeta\) with \(\mathcal{O}\) is the set of observables of Pauli string type, such that the shadow estimator \(\widehat{\rho}\) constructed using unitaries \(U_i\) uniformly sampled from \(\zeta'\), given by 
\begin{equation}
    \widehat{\rho} = \mathbbm{E}_{i, \hat{k}} \left[ \mathcal{M}_p^{-1} \left( U_i^\dagger \proj{\hat{k}} U_i \right) \right].
    \label{eq:estimator}
\end{equation}  
satisfies the following relation in the limit of large measurements:
\begin{equation}
    \expv{{O}}{\rho} = \Tr({O} \widehat{\rho}).
    \label{eqn:ObS}
\end{equation}  
for some non-trivial set of Pauli string observables $O\in\mathcal{O}$.

To illustrate this idea, we consider a two-qubit system \( \rho \), where we sample specific subsets of unitaries from $\zeta=\mathrm{Cl}(2)^{\otimes 2}$ and construct \( \widehat{\rho} \) using the pseudo-inverse map \(\mathcal{M}^{-1}_p(\cdot)\). At first, we take the unitary set consisting of only one unitary \(\zeta'=\{\mathbb{1}\otimes\mathbb{1}\}\), we generate the shadow estimator, which in the limit of a large number of samples reduces to 
\begin{align}
&\widehat{\rho}_{\mathbb{1}\otimes\mathbb{1}} =p~\text{diag}(\rho)-\mathbb{1},~~\text{where}\nonumber\\
&~~~~~~~~~\text{diag}(\rho)=(\rho_{00,00}~~\rho_{01,01}~~\rho_{10,10}~~\rho_{11,11}) 
\end{align}
In general, it does not estimate any non-trivial Pauli string observable except \(\mathbb{1}\otimes\mathbb{1}\) when \(p=5\).
None of the other density matrix elements can be recovered. Consequently, additional post-processing is required to estimate observables such as \(
\mathbb{1} \otimes Z, ~ Z \otimes \mathbb{1}, ~ Z \otimes Z
\).
Now we explore other cases: \(\zeta'=\{H\otimes H\}\) and \(\zeta'=\{HS\otimes HS\}\), for which the estimator in Eq.~\eqref{eq:estimator} approaches \(\widehat{\rho}_{H\otimes H}\) and \(\widehat{\rho}_{HS\otimes HS}\) respectively, given by
\begin{align}
\widehat{\rho}_{H\otimes H}&= -\mathbb{1}+\frac{p}{4}\mathcal{B}_H.\\
\widehat{\rho}_{HS\otimes HS}&= -\mathbb{1}+\frac{p}{4}\mathcal{B}_{HS}.
\end{align}
The explicit form of \(\mathcal{B}_H\) and \(\mathcal{B}_{HS}\) matrices are given in the Appendix \ref{2not} in Eqns. \eqref{eq:HXH} and \eqref{eq:HSXHS} respectively. The estimator $\widehat{\rho}_{H\otimes H}$ accurately estimates 
$\{X\otimes\mathbb{1},~ \mathbb{1}\otimes X,~ X\otimes X\}$, and 
$\widehat{\rho}_{HS\otimes HS}$ estimates 
$\{Y\otimes\mathbb{1},~ \mathbb{1}\otimes Y,~ Y\otimes Y\}$ when $p=1$, but neither estimator captures any non-trivial combinations of the above-mentioned Pauli strings for $ p\neq 1$.
However, an explicit scan through all the subsets suggests that a convenient set of unitaries that satisfy the above requirements can be found to be
\begin{itemize}
    \item[] \(\zeta_X = \{\mathbb{1}\otimes\mathbb{1},H\otimes H,H\otimes HS,HS\otimes H,HS\otimes HS\}\),
    \item[] \(\zeta_1 = \{\mathbb{1}\otimes\mathbb{1},H\otimes\mathbb{1},\mathbb{1}\otimes H,\mathbb{1}\otimes HS,HS\otimes\mathbb{1}\}\).
\end{itemize}
The estimator Eq.~\eqref{eq:estimator} with $p=5$, and unitaries sampled from \(\zeta_X\) and \(\zeta_1\) approach \(\widehat{\rho}_X\) and \(\widehat{\rho}_1\) respectively and are given by:
\begin{equation}
\resizebox{\linewidth}{!}{$
\widehat{\rho}_X=
\renewcommand{\arraystretch}{1.5}
\begin{pmatrix}
    \rho_{00,00} & \rho_{00,01}+\rho_{10,11} & \rho_{00,10}+\rho_{01,11} & \rho_{11,11} \\
    \rho_{01,00}+\rho_{11,10} & \rho_{01,01} & \rho_{01,10} & \rho_{01,11}+\rho_{00,10} \\
    \rho_{10,00}+\rho_{11,01} & \rho_{10,01} & \rho_{10,10} & \rho_{10,11}+\rho_{00,01} \\
    \rho_{00,11} & \rho_{11,01}+\rho_{10,00} & \rho_{11,10} + \rho_{01,00} & \rho_{11,11}  
\end{pmatrix}
$}
\label{eq:2X}
\end{equation}
\begin{equation}
\resizebox{\linewidth}{!}{$
\widehat{\rho}_1=
\renewcommand{\arraystretch}{1.5}
\begin{pmatrix}
    2\rho_{00,00}-\rho_{11,11} & \rho_{00,01} & \rho_{00,10} & 0 \\
    \rho_{01,00} & 2\rho_{01,01}-\rho_{10,10} & 0 & \rho_{01,11} \\
    \rho_{10,00} & 0 & 2\rho_{10,10}-\rho_{01,01} & \rho_{10,11} \\
    0 & \rho_{11,01} & \rho_{11,10}  & 2\rho_{11,11}-\rho_{00,00}  
\end{pmatrix}
$}.
\label{eq:2NX}
\end{equation}
The estimator \(\widehat{\rho}_X\) accurately captures the diagonal and anti-diagonal elements of the density matrix. The other remaining off-diagonal elements are captured in their respective positions via the estimator \(\widehat{\rho}_1\) (see Fig. \ref{fig:partial shadow} (b)).

Note that \(\widehat{\rho}_X\) allows to calculate, without any additional processing, the expectation values of Pauli strings \(~\mathcal{O}_X=\{\mathbb{1}\otimes\mathbb{1},~\mathbb{1} \otimes Z, ~ Z \otimes \mathbb{1}, ~ Z \otimes Z, ~X \otimes X,~Y \otimes Y,~X \otimes Y,~X \otimes Y\}\) and their arbitrary linear combinations. These operators set includes widely studied models such as the  XYZ Hamiltonian with longitudinal field and therefore can be useful in efficient use of shadow tomography approaches for variational quantum algorithms on such Hamiltonians. On the other hand, the estimator \(\widehat{\rho}_1\) allows one to calculate the expectation values of the Pauli strings
\(~\mathcal{O}_1=\{\mathbb{1}\otimes X,~\mathbb{1} \otimes Y, ~ X \otimes \mathbb{1}, ~ Y \otimes \mathbb{1}, ~X \otimes Z,~Y \otimes Z,~Z \otimes X,~Z \otimes X\}\)
and their arbitrary linear combinations via Eq.~\eqref{eqn:ObS}. Combining the two estimates allows complete characterization of the density matrix. More importantly, if the observables to be estimated are in either of the sets, estimation can be performed more efficiently than full shadow tomography.

The structure of $\zeta_X$ and $\zeta_1$ can be generalized by tuning $p-$values to get similar efficient estimates of larger Pauli strings inside an $n$-qubit system.
In this paper, PQST is formulated to estimate a restricted class of density matrix elements, referred to as active orders (See Sec.~\ref{Sec:PQST}). All elements with the same active order can be simultaneously estimated using a simple pseudo-inverse map (Sec~\ref{generalprotocol}), which can be implemented efficiently, without additional computational overhead.

Lastly, we note that the unitary set \(\zeta=\{\mathbb{1}, H, HS\}^{\otimes n}\) is also a subset of Pauli basis measurements \(\zeta\subset\text{Cl}(2)^{\otimes n}\). However, we find that when the channel's inverse for the case of Pauli basis measurements given in the Eq.~\eqref{eqn:pauliinv} is used in conjugation with $\zeta_X$ or \(\zeta_1\), we do not recover correct matrix elements without using correction factors.

\subsection{Partial Quantum Shadow Tomography (PQST)}
\label{Sec:PQST}
The PQST protocol involves performing independent shadow tomography using different sets of unitaries \(\zeta_1, \zeta_2, \dots\), where each set \(\zeta_i\) is associated with a pseudo-inverse described in Eq.~\eqref{eqn:Inverse} for some appropriate strength $p_1,p_2,\dots$, yielding Partial Shadow Estimators (PSEs) \(\widehat{\rho}_1, \widehat{\rho}_2, \dots\) 
\begin{align}
    \zeta_1, p_1 : \rho &\xrightarrow{QST} \widehat{\rho}_1
    \nonumber \\
    \zeta_2,p_2 : \rho &\xrightarrow{QST} \widehat{\rho}_2
    \nonumber \\
    \zeta_3,p_3 : \rho &\xrightarrow{QST} \widehat{\rho}_3
    \\
    &\vdots \nonumber
\end{align}
Each PSE captures partial disjoint pieces of information about the quantum state. As shown in Sec. \ref{generalprotocol}, these PSEs collectively reconstruct the full density matrix by:
\begin{equation}
\rho = \sum_{i=1}^N \mathcal{P}_i(\widehat{\rho}_i),
\label{eq:rhorecover}
\end{equation}
where \( P_i(\cdot) \) projects the density matrix elements, preserving those elements that each estimator \( \widehat{\rho}_i \) can estimate. This process is illustrated in Fig. \ref{fig:partial shadow}.\\
Here, we introduce the active notation for density matrix elements~\cite{levitt2008spin}. Let \(\rho\) be the density matrix of an \(n\)-qubit quantum system, expressed in the computational basis \(\{|k\rangle\}_{k=0}^{2^n-1}\). The matrix element  \(\rho_{ij} = \langle i | \rho | j \rangle\) corresponds to the transition amplitude between the basis states \(|i\rangle\) and \(|j\rangle\). Here, \(i\) and \(j\) are bit strings of length \(n\), representing the computational basis states \(|i\rangle\) and \(|j\rangle\) of the \(n\)-qubit system, given by the tensor product of single-qubit basis states.
A matrix element \(\rho_{ij}\) is said to be {\(d\)-active} if the bit strings \(i\) and \(j\) differ in exactly \(d\) sites. This difference is quantified by the Hamming distance, which counts the number of positions where \(i\) and \(j\) have different bits.


\subsection{PQST in ensemble systems \label{sec:pqstensemble}}
In the context of ensemble systems like in NMR, PQST can be
implemented efficiently by measuring population elements via diagonal tomography, which is equivalent to performing a large number of  projective measurements
\cite{DiagonalTomo,LEE2002349, AAST}. In this case, averaging over computational basis states \(\ket{k}\) is captured by diagonal tomography.
Given a quantum state $\rho$ encoded in an ensemble quantum processor, the PQST is realized via the following steps:
\begin{itemize}
\item[(i)] First rotate the target state $\rho$ under $U_i\in \zeta$, i.e.,
\( \rho \rightarrow U_i \rho U_i^\dagger \).
\item[(ii)] Readout all the diagonal elements \( \dmel{k}{U_i \rho U_i^\dagger} = P_{ik}\) by performing diagonal tomography in the computational basis $\{\ket{k}\}$ .
 \item[(iii)] Reverse rotate the diagonal state
 on a classical processor to obtain the given matrix
 \begin{align}
 \sum_k U_i^\dagger \left( P_{ik} \proj{k} \right) U_i 
 \end{align}
\item[(iv)] Invert using the pseudo-inverse map of Eq. \eqref{eqn:Inverse} and average over the choices of unitaries to construct the PSE
\begin{equation}   
\widehat{\rho}_\zeta = \mathbbm{E}_{U_i\in \zeta} \left[ {\mathcal M}^{-1}_{p}\left(\sum_k U_i^\dagger \left( P_{ik} \proj{k} \right) U_i \right)\right].
\label{eqn:Ensemble_case}
\end{equation} 
\end{itemize}
PQST is a general protocol adapted for near-term quantum devices. In platforms like ion traps or superconducting qubits, which use single-copy projective measurements, it aligns closely with classical shadow tomography. In contrast, ensemble-based systems such as NMR~\cite{LEE2002349} enable exact estimation of density matrix elements via ensemble averaging, effectively performing partial quantum state tomography.
Thus, PQST serves as an efficient and scalable alternative to full or partial state tomography in ensemble-based quantum systems.

In the following, we consider PQST for quantum registers of different sizes.

\subsection{Shadow protocol for 1-qubit system
\label{sec:1qpqst}}
Using the uniform sampling of unitaries from set $\zeta = \{\mathbb{1}, H, HS\}$, the full shadow estimator can be written as 
\begin{equation}
    \widehat{\rho}= \mathbbm{E}_{\hat{k},U_i\in \zeta}\left[\mathcal{M}_3^{-1} \left( U_i^\dagger \proj{\hat{k}}  U_i\right)\right], 
\end{equation}
where $\hat{\ket{k}}$ is the measurement outcome in the computational basis $\{\ket{k}\}$ after rotating \(\rho\) with \(U_i\). It requires only three unitaries for the full density matrix estimation without any approximation in the large measurement limit. Here, we used the strength parameter $p=3$ for the pseudo-inverse map given in Eq.~\eqref{eqn:Inverse}.
\subsection{PQST for a 2-qubit system
\label{sec:2qpqst}}
\begin{table}
    \centering
    \renewcommand{\arraystretch}{2}
    \setlength{\tabcolsep}{11pt}
    \begin{tabular}{|c|c|c|c|c|}
        \hline
        {} & \(\bra{00}\) & \(\bra{01}\) & \(\bra{10}\) & \(\bra{11}\) \\ \hline
        \(\ket{00}\) & \cellcolor{gray!25}\(\rho_{00,00}\) & \(\rho_{00,01}\) & \(\rho_{00,10}\) & \cellcolor{gray!25}\(\rho_{00,11}\) \\ \hline
        \(\ket{01}\) & \(\rho_{01,00}\) & \cellcolor{gray!25}\(\rho_{01,01}\) & \cellcolor{gray!25}\(\rho_{01,10}\) & \(\rho_{01,11}\) \\ \hline
        \(\ket{10}\) & \(\rho_{10,00}\) & \cellcolor{gray!25}\(\rho_{10,01}\) & \cellcolor{gray!25}\(\rho_{10,10}\) & \(\rho_{10,11}\) \\ \hline
        \(\ket{11}\) & \cellcolor{gray!25}\(\rho_{11,00}\) & \(\rho_{11,01}\) & \(\rho_{11,10}\) & \cellcolor{gray!25}\(\rho_{11,11}\) \\ \hline
    \end{tabular}
    \caption{Classification and estimation of active orders. 
    Shaded entries correspond to 0-active and 2-active terms 
    estimated by \(\widehat{\rho}_X\). 
    The remaining off-diagonal elements represent 1-active terms, estimated by 
    \(\widehat{\rho}_1\). 
    A detailed channel description is provided in Appendix~\ref{2not}.}
    \label{Tabel1}
\end{table}
For the 2-qubit case, the full tomography complete set consists of all tensor products of the single qubit unitaries, i.e., \(\{\mathbb{1}, H, HS\}^{\otimes 2}\), which has 9 unitaries. We can divide these unitaries into two sets (see Fig. \ref{fig:partial shadow} (b)).\\

 \(\zeta_X = \{\mathbb{1}\otimes\mathbb{1},H\otimes H,H\otimes HS,HS\otimes H,HS\otimes HS\}\).\\

The shadow estimator $\widehat{\rho}_X$ generated from \(\zeta_X\) with $p=5$, estimates zero-active (diagonal) and two-active (anti-diagonal) elements of a 2-qubit density matrix.\\

 \(\zeta_1 = \{\mathbb{1}\otimes\mathbb{1},H\otimes\mathbb{1},\mathbb{1}\otimes H,\mathbb{1}\otimes HS,HS\otimes\mathbb{1}\}\).\\

The shadow estimator $\widehat{\rho}_1$ generated from \(\zeta_1\) with $p=5$, estimates all the single-active elements of a 2-qubit density matrix.
Further $\zeta_1=\zeta_{1a}\cup\zeta_{1b}$ can be further separated into two subsets.
\begin{enumerate}
    \item[(i)] \(\zeta_{1a} = \{\mathbb{1}\otimes\mathbb{1},H\otimes\mathbb{1},HS \otimes\mathbb{1}\}\) with $p=3$, yields PSE \(\widehat{\rho}_{1a} =\{ \rho_{00,10},\rho_{01,11}, \rho_{10,00}, \rho_{11,01} \}\) estimating single-active terms of the first qubit.
    \item[(ii)] \(\zeta_{1b} = \{\mathbb{1}\otimes\mathbb{1},\mathbb{1}\otimes H,\mathbb{1}\otimes HS\}\) with $p=3$, yields PSE \(\widehat{\rho}_{1b} =\{ \rho_{00,01},\rho_{01,00}, \rho_{10,11}, \rho_{11,10} \} \) estimating single-active terms of the second qubit.
\end{enumerate}
Tab.~\ref{Tabel1} captures the active orders which are efficiently estimated by different PSEs.
\subsection{PQST for a 3-qubit system}
\label{sec:3sets}
\begin{figure}
    \centering
    \includegraphics[width=0.7\linewidth, clip=true, trim={6cm, 12.5cm, 6.2cm, 4.2cm}]{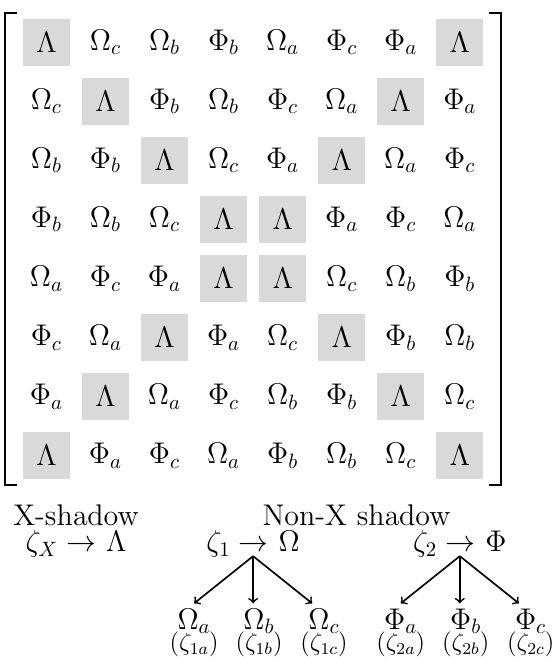}
    \caption{
    PQST of a 3-qubit system.
    The estimator \( \widehat{\rho}_X \), generated by \( \zeta_X \), efficiently estimates the density matrix elements corresponding to the \( \Lambda \) positions which constitutes the X-shadow, while the estimators \( \widehat{\rho}_1 \) and \( \widehat{\rho}_2 \), generated by \( \zeta_1 \) and \( \zeta_2 \), respectively, efficiently estimate the single-active terms \( \Omega\) and double-active terms \( \Phi \). \( \Omega\) and \(\Phi\) can be further divided into subsets as mentioned in Sec. \ref{sec:3sets}.
    }
    \label{fig:Xshadow}
\end{figure}

The full unitary set \(\{\mathbb{1}, H, HS\}^{\otimes 3}\) can be divided into three sets (Fig.~\ref{fig:Xshadow}) 
\begin{itemize}
    \item [(i)]The unitary set \( \zeta_X = \{\mathbb{1}^{\otimes 3}, u_1\otimes u_2 \otimes u_3\} \), where \( u_i \in \{H, HS\} \), comprising $1 + 2^3 = 9$ three-qubit unitaries that apply either $\mathbb{1}$ to all qubits or a non-identity unitary ($H$ or $HS$) to each qubit. Uniform sampling from $\zeta_X$ yields the PSE $\widehat{\rho}_X$, which estimates the diagonal ($0$-active) and anti-diagonal ($3$-active) components of $\rho$ (denoted $\Lambda$ in Fig.~\ref{fig:Xshadow}) via the pseudo-inverse transformation in Eq.~\eqref{eqn:Inverse} with strength parameter $p=9$.

    \item [(ii)]\(\zeta_1=\{\mathbb{1}^{\otimes 3},~u_1\otimes\mathbb{1}\otimes\mathbb{1}, \mathbb{1}\otimes u_2\otimes\mathbb{1}, \mathbb{1}\otimes\mathbb{1}\otimes u_3\}\) consisting of $7$ unitaries in which a non-identity operation ($H$ or $HS$) acts on exactly one qubit while the remaining qubits are left unchanged. Uniform sampling from $\zeta_1$ yields the PSE $\widehat{\rho}_1$, which estimates the single-active components of $\rho$ (denoted $\Omega$ in Fig.~\ref{fig:Xshadow}) via the pseudo-inverse transformation, Eq.~\eqref{eqn:Inverse} with strength parameter $p=7$.  It can be further divided into the following subsets:
    \begin{align}
    \zeta_{1a} &= \{\mathbb{1}^{\otimes 3},\, u_1 \otimes \mathbb{1} \otimes \mathbb{1}\}, \quad u_1 \in \{H, HS\}, \\
    \zeta_{1b} &= \{\mathbb{1}^{\otimes 3},\, \mathbb{1} \otimes u_2 \otimes \mathbb{1}\}, \quad u_2 \in \{H, HS\}, \\
    \zeta_{1c} &= \{\mathbb{1}^{\otimes 3},\, \mathbb{1} \otimes \mathbb{1} \otimes u_3\}, \quad u_3 \in \{H, HS\},
    \end{align}
    each containing three unitaries used to construct the PSEs $\widehat{\rho}_{1a}$, $\widehat{\rho}_{1b}$, and $\widehat{\rho}_{1c}$, respectively. 
    These estimators estimate the single-active components of $\rho$ associated with the first, second, and third qubits (denoted $\Omega_a$, $\Omega_b$, and $\Omega_c$ in Fig.~\ref{fig:Xshadow}) via the pseudo-inverse transformation in Eq.~\eqref{eqn:Inverse} with strength parameter $p = 3$.\\
    Note that each subset $\zeta_{1i}$ contains three unitaries. 
    The terms reconstructed from $\zeta_{1i}$ correspond to the single-active components of the qubit on which the applied unitary is non-identity. 
    To estimate the single-active terms of the first and second qubits jointly, we take the union \(\zeta_{1ab} = \zeta_{1a} \cup \zeta_{1b},\) which contains five unitaries, and perform reconstruction using the pseudo-inverse transformation in Eq.~\eqref{eqn:Inverse} with strength parameter $p = 5$. 
    This construction naturally extends to other combinations of the subsets.

    \item[(iii)] \(\zeta_2=\{\mathbb{1}^{\otimes 3},~u_1\otimes u_2\otimes\mathbb{1}, \mathbb{1}\otimes u_2\otimes u_3,u_1\otimes\mathbb{1}\otimes u_3\}\) consisting of $13$ unitaries in which two qubits are acted upon by non-identity operations ($H$ or $HS$) while the remaining qubit is left unchanged. 
    Uniform sampling from $\zeta_2$ yields the PSE $\widehat{\rho}_2$, which estimates the two-active components of $\rho$ (denoted $\Phi$ in Fig.~\ref{fig:Xshadow}) via the pseudo-inverse transformation in Eq.~\eqref{eqn:Inverse} with strength parameter $p = 13$. It can be further divided into the following subsets:
    \begin{align}
    \zeta_{2a} &= \{\mathbb{1}^{\otimes 3},\, u_1 \otimes u_2 \otimes \mathbb{1}\},~ u_1, u_2 \in \{H, HS\}, \\
    \zeta_{2b} &= \{\mathbb{1}^{\otimes 3},\, \mathbb{1} \otimes u_2 \otimes u_3\},~ u_2, u_3 \in \{H, HS\}, \\
    \zeta_{2c} &= \{\mathbb{1}^{\otimes 3},\, u_1 \otimes \mathbb{1} \otimes u_3\},~ u_1, u_3 \in \{H, HS\},
    \end{align}
    each containing five unitaries used to construct the PSEs $\widehat{\rho}_{2a}$, $\widehat{\rho}_{2b}$, and $\widehat{\rho}_{2c}$, respectively. 
    These estimators estimate the two-active components of $\rho$ associated with the qubit pairs $(1,2)$, $(2,3)$, and $(1,3)$ (denoted $\Phi_a$, $\Phi_b$, and $\Phi_c$ in Fig.~\ref{fig:Xshadow}) via the pseudo-inverse transformation in Eq.~\eqref{eqn:Inverse} with strength parameter $p = 5$.\\
    Each of these subsets $\zeta_{2i}$ consists of five unitaries with $p=5$. To enable the simultaneous estimation of 2-active terms where qubits $(1,2)$ and $(2,3)$ are active, we define $\zeta_{2ab} = \zeta_{2a} \cup \zeta_{2b}$, which contains nine unitaries. The estimator is then constructed by uniform sampling of unitaries from $\zeta_{2ab}$ with pseudo-inverse map strength $p=9$ in Eq.~\eqref{eqn:Inverse}. This allows us to estimate 2-active terms where $(1,2)$ and $(2,3)$ qubits are active.
\end{itemize}
Through these examples, it becomes evident that the strength of the pseudo-inverse map is determined by the cardinality of the corresponding unitary set \(\zeta\), given by $p=|\zeta|$. This enables the generation of PSEs Eq.~\eqref{eq:estimator}, which facilitates the estimation of density matrix elements corresponding to specific active orders.
\section{Generalized PQST Protocol
\label{generalprotocol}}
For an \(n\)-qubit system, the full unitary set is given by \(3^n\) unitary operations \(\zeta = \{\mathbb{1}, H, HS\}^{\otimes n}.\)
We consider the problem of estimating 
the $A$-active matrix elements of $\rho$, where only qubits in the subset $A \subseteq \{1, \dots, n\}$ of all qubits are active.  
To achieve this, we introduce a set of unitaries $\zeta_A$, which consists of the identity operator \( \mathbb{1}^{\otimes n} \) along with unitaries that act trivially on the complement of \( A \) qubits and as non-trivial operations \( u \in \{H, HS\} \) on the qubits within \( A \). The unitary set $\zeta_A$ enables the estimation of all $A$-active terms of the density matrix using the estimator in Eq.~\eqref{eq:estimator} and the pseudo-inverse map Eq.~\eqref{eqn:Inverse} with \(p = |\zeta_A| = 2^{|A|} + 1,\) where \(|.|\) denotes the cardinality.
We find that the idea can be extended to simultaneously calculate all $A$- active and $B$- active density matrix elements for two subsets $A$ and $B$ of the same cardinality. For this we use the set $\zeta_{A}\cup \zeta_{B}$ and use the estimator in Eq.~\eqref{eq:estimator} with the pseudo-inverse map with $p=|\zeta_{A}\cup \zeta_{B}|$. This can be generalized to a combination of more subsets $A_1,A_2\dots$ all of the same size. For instance all the $m$-active elements ($1\leq m\leq n$) of $\rho$ can be calculated using a unitary set of size $p=\binom{n}{m}\times 2^{m} + 1$. To be more precise, we have outlined the protocol for an $n$-qubit system in Appendix~\ref{GP}.

\section{PQST for structured operators}
\label{Sec:StrOPr}
We present PQST for structured operators, particularly for specific density matrices or observable structures, for which PQST enables efficient estimation.  
\subsection{X Shadow Tomography}
\label{Sec:Xtomo}
If the observable has an X-structure (containing only diagonal and anti-diagonal entries), its expectation value can be determined using the X-shadow for any density matrix. Conversely, if the density matrix itself has an X-structure, the expectation value can be computed using the X-shadow for any arbitrary observable.
\paragraph*{X-structured operators}: We can use the X-shadow which samples unitaries uniformly from \(\zeta_X\) to compute the expectation of observables of the form given by 
\begin{align}
    \mathcal{P} = \mathcal{P}_{Z}+\mathcal{P}_{XY},
    \label{eq:Xstructure}
\end{align}
where $\mathcal{P}_Z$ and $\mathcal{P}_{XY}$ can be any linear combinations of Pauli strings made of $\{\mathbb{1},Z\}$ and $\{X,Y\}$  operators respectively.
Note that ${\mathcal P}$ is an X-structured operator, i.e., it contains only diagonal and anti-diagonal elements in the computational basis for the $n$ qubits.
Many Hamiltonians of commonly studied systems, such as the transverse field Ising model, as well as $XXZ$ and $XYZ$ models with longitudinal field, contain X-structured operators for every adjacent pair of qubits. Thus, their expectation values can be estimated efficiently using the X-shadow on every adjacent pair. This reduces the cost of unitary sampling while still capturing the relevant correlations in $XX$, $YY$, and $XY$ interactions.

In cases where the observable ${\mathcal O}$ does not have the X-structure, we can still estimate the expectation of such observables via a unitary transformation \(\mathcal{U}^\dagger\), which maps the observable \(\mathcal{O}\) to an observable of the form \(\mathcal{P}\) in Eq.~\eqref{eq:Xstructure}. The expectation value of $\mathcal{O}$ can then be calculated as the expectation value of the X-structured operator $\mathcal{P}$ of the rotated state $\mathcal{U}\rho \mathcal{U}^\dagger$. 
\begin{align}
    \mathcal{O}&= \mathcal{U}^\dagger\mathcal{P}\mathcal{U}.
    \label{Transform}
\end{align}
Single-qubit unitaries can be implemented efficiently and with high fidelities, enlarging the set of operators whose expectation values can be estimated using X-shadow tomography. For example, 2-qubit observables \(Z\otimes X,~Z\otimes Y\) are not directly accessible through the X-shadow. They can be estimated by employing X-shadow tomography on \(\mathcal{U}\rho \mathcal{U}^\dagger\) state, where \(\mathcal{U}=\mathbb{1}\otimes H,~\mathcal{U}= \mathbb{1}\otimes HSH\) respectively, effectively calculating expectation of \(Z\otimes Z\) w.r.t. the rotated state in each case. \\
In the case of a larger number of qubits, the above statements generalize.\\
1) There are specific groups of observables, \(\mathcal{A}_1, \mathcal{A}_2, \dots\), 
that can be determined directly by sampling from certain measurement sets 
\(\zeta_1, \zeta_2, \dots\).\\ 
2) If the observable \(O\) 
of interest does not belong to any \(\mathcal{A}_i\), we can apply a suitable 
unitary transformation so that \(O\) is mapped into one of these groups. Then, 
by performing partial tomography on the inverse-rotated state, allow us to learn  \(\langle O \rangle\).
\paragraph*{X-structured density matrices}: X-tomography can also be used to estimate the expectation values of arbitrary operators on states whose density matrix is known to be X-structured. Such states with X-structured density matrices include Werner states and Bell diagonal states (convex sums of Bell states), etc.

\subsection{Non-X shadow tomography}
In cases where the state or observable does not have an X-structure, we can sample unitaries from smaller subsets $\{\zeta_d\}$ to extract relevant elements to estimate certain observables. For example, if we need to estimate a combination of terms like \(\Phi_a\) and \(\Omega_b\). In that case, we can sample from subsets \(\zeta_{2a}\) and \(\zeta_{1b}\), respectively and construct the PSEs \(\widehat{\rho}_{2a}\) and \(\widehat{\rho}_{1b}\) separately and estimate all the \(\Phi_a\) and \(\Omega_b\) terms.
Since this approach samples unitaries from a much smaller subset, it is advantageous compared to performing full shadow tomography.

\section{Sample Complexity and error estimation}
\label{Sec:Sample}
In this section, we argue that with the choice of $\mathcal{M}^{-1}$ and the restricted set of unitaries, the PQST protocol gives an unbiased estimator for the structured operators. We also present a bound on the shadow norm, which can in turn be used to place upper bounds on the variance of the estimator, error in their mean, and sampling complexity. Sample complexity is particularly relevant for single-copy, projective measurement-based systems, in contrast to ensemble-based architectures, which provide exact expectation values. Numerical results in subsequent sections validate the analysis presented here.
\subsection{X-tomography}
We sample unitaries from the subset $\mathcal{U}=\zeta_X$ to perform X-shadow tomography:
\begin{align}
    \zeta_X=  \{\mathbb{1}^{\otimes n}\} \cup \left\{ \bigotimes_{i=1}^{n} u_i \;\middle|\; u_i \in \{H, HS\} \right\},
\end{align}
which contains $2^n+1$ unitaries. The unknown density matrix to be estimated has an expansion $\rho=\sum_{\vec{\mu}} c_{\vec{\mu}} \sigma_{\vec{\mu}}$ where $\sigma_{\vec{\mu}}$ is a Pauli string  $\sigma_{\mu_1}\sigma_{\mu_2}\dots\sigma_{\mu_n}$ and $c_{\vec\mu}$ is a real number (We use $\sigma_{0,1,2,3}$ to be the matrices $\mathbb{1},X,Y,Z$ here). 
Terms in the expansion for $\rho$ which contain only $\mathbb{1},Z$ or only $X,Y$ fully determine the X-structured operator expectation values. We define the span of $\{\mathbb{1},Z\}^{\otimes n}$ and span of $\{X,Y\}^{\otimes n}$ as $V_d$ and $V_{ad}$; and define $V_\times =V_d\oplus V_{ad}$.
We claim that the expectation value of an X-structured operator can be obtained from the unbiased estimator
\begin{equation}
\hat{O} = \mathrm {Tr}\left [O \mathcal{M}^{-1}[U\vert b \rangle \langle  b \vert U^\dagger]\right ],
\label{eq:Oestimator}
\end{equation}
where $\mathcal{M}^{-1}(A)=pA-{\mathrm{Tr}}(A)\mathbb{1}$ where $p=2^n+1$.
To see this, we note two points. Firstly, the expectation $\mathbbm{E}_{U,b}[\hat{O}]$ can be written as 
\begin{equation}
\frac{1}{p}\sum_{U\sim\mathcal{U}}\sum_{b}\langle b \vert \left(pUOU^{\dagger}-\mathrm {Tr}(O)\mathbb{1}\right) \vert b \rangle \left\langle b|U\rho U^{\dagger}|b\right\rangle,\label{eq:expectOfestimator}
\end{equation}
where we used the Hermiticity of $\mathcal{M}^{-1}$ in the operator space with Hilbert Schmidt norm.
Secondly, for each non-identity Pauli string $\sigma_{\mu}\in V_\times$ there is exactly one $U\in \mathcal{U}$ such that $U\sigma_{\mu}U^\dagger\in V_d$. For an X structured operator $O=\sum_{\vec{\mu} \in V_\times} a_{\vec \mu} \sigma_{\vec \mu}$ and a general density matrix $\rho=\sum_{\vec{\mu}} c_{\vec \mu} \sigma_{\vec \mu}$, (See Appendix~\ref{Sec:ExpectationCal}) and using the above mentioned properties of $\mathcal{U}$, we get that $\mathbbm{E}_{U,b}[\hat{O}]=2^n\sum_{\vec \mu} {a_\mu}c_\mu = \mathrm {Tr}(\rho O)$.
Having shown that $\hat{O}$ is an unbiased estimator of $\langle O \rangle$, we proceed to analyse the variance
From Ref.~\cite{Huang2020} (Lemma S1), the variance is given by
\begin{align}
 \mathrm{Var}[\hat{O}] 
= \mathbbm{E}\!\left(\hat{O} - \mathbbm{E}[\hat{O}] \right)^{2} 
\leq \left\|O- \frac{\mathrm{tr}\!\left(O\right)}{2^{n}}\mathbb{1}\right\|^{2}_\text{shadow}.
\end{align}
The shadow norm $\|O\|_{\text{shadow}}^2$ is defined as the following 
\begin{multline}
 \max_{\sigma:\text{state}}
 \mathbbm{E}_{U\sim \mathcal{U}}  \sum_{b \in \{0,1\}^n} 
\bra{b} U \sigma U^\dagger \ket{b}
\bra{b} U \mathcal{M}^{-1}(O) U^\dagger \ket{b}^2. 
\end{multline}
For an X-structure observable \(O=O_{ad}+O_d\), (where $O_{ad},~O_d$ are anti-diagonal, diagonal) and unitary ensemble \(\mathcal{U}=\zeta_X\) the norm is bound by 
\begin{align}
\left\|O-\frac{\mathrm{Tr}(O)}{2^n}\mathbb{1}\right\|^2_{\mathrm{shadow}} \leq &\left(1+\frac{1}{2^n}\right)(\|O_{ad}\|^2_\mathrm{HS}+2^n \|O_{d}\|^2_{\infty})\nonumber\\
&\leq 2(2^n+1)\|O\|_\infty^2,
\label{eq.X}
\end{align}
where \(\|A\|_\mathrm{HS}\) and \(\|A\|_{\infty}\) denote the Hilbert--Schmidt and operator norms, respectively (see Appendix~\ref{Sec:Proof} for details). The upper bound derived for PQST is strictly lower than the corresponding bounds for Pauli measurements. The variance determined by the shadow norm yields the standard error of the mean, which is bounded by \(\|O\|_{\mathrm{shadow}}/\sqrt{N}\) for \(N\) estimator evaluations. 

Theorem~1 of Ref.~\cite{Huang2020} relates the sampling complexity of the median-of-means estimator to the shadow norm; a reduced shadow norm directly implies improved sampling complexity. The lower shadow norm for PQST compared to the full Pauli measurement scheme suggests lower sampling complexity for the PQST protocol. 
In particular, for purely anti-diagonal observables \((O=O_{ad})\), the shadow norm is bounded by the Hilbert--Schmidt norm with a smaller pre-factor than in the Clifford measurement setting, thereby reducing the sampling complexity for such \(n\)-active observables.
\subsection{Non-X tomography}
\label{Sec:SampleNX}
Before proceeding, we define $k$-active observables.
For any Pauli string, \(P\in \mathcal{P}_n = \{\mathbb{1},X,Y,Z\}^{\otimes n}\),   
we can define its active order as the total number of $X$s and $Y$s in the string. We can define the space $V_k$ of the $k$-active operators as the span over reals of all $k$-active Pauli strings. For each Pauli string $\sigma_{\vec \mu}$ in this space, we can define $A(\vec{\mu})$ as the set of sites in which $\sigma_{\vec\mu}$ has $X$ or $Y$ operators.
Observables $O_k=\sum_{\vec\mu\in V_k} a_{\vec \mu} \sigma_{\vec \mu}$ in this space are called {$k$-active observables}.

If $O$ is a $k$-active observable that is a linear combination of $k$-active Pauli strings from a set $S\subset V_k$, then the expectation value can be obtained by sampling from the unitary set
\begin{align}
\zeta_k= 
\bigcup_{\vec{\mu} \in S} \Bigg\{ 
    \bigotimes_{i=1}^n u_{i}, \;
    & u_i \in \{H,HS\} \;\; \mathrm{if }~ i \in A(\vec{\mu}),\nonumber \\
    &u_i = \mathbb{1} \;\; \mathrm{otherwise} 
\Bigg\}
\;\cup\; \{ \mathbb{1}^{\otimes n} \},
\label{eq.zetak}
\end{align}
and using the pseudo-inverse map \(\mathcal{M}^{-1}(A)=pA-\mathrm{Tr}(A)\mathbb{1}\) where \(p=|\mathcal{U}|\). Eq.~\eqref{eq:Oestimator} is an unbiased estimator of the expectation value and has a variance (See Appendices~\ref{NX-observable} and~\ref{Sec:Proof}). For non-X tomography, the shadow norm satisfies 
\(\|O\|_\mathrm{shadow}=\sqrt{\tfrac{p}{2^n}}\,\|O\|_\mathrm{HS}\), 
which directly yields the variance bound
\begin{equation}
{\mathrm{Var}}[\hat{O}_k]\leq\|O_k\|_\mathrm{shadow}^2\implies {\mathrm{Var}}[\hat{O}_k]\leq{\frac{p}{2^n}}\|O_k\|_\mathrm{HS}^2.
\end{equation}
While this presents a bound which is similar in scaling to the variance in the case of Clifford unitaries, explicit numerical examples presented in the next section suggest that, for structured observables, partial tomography is more efficient.
Another important point is that, for estimating a $k$-active observable $O_k$, it is not necessary to sample $\mathbb{1}^{\otimes n}$ from the set $\zeta_k$. However, including $\mathbb{1}^{\otimes n}$ enables the estimation of certain diagonal observables, albeit with a slightly larger error bound (see Appendix~\ref{Sec:NXcal}).

\begin{figure*}
    \includegraphics[width=17.5cm, clip=true, trim={0.4cm 0.2cm 0.2cm 0.cm}]{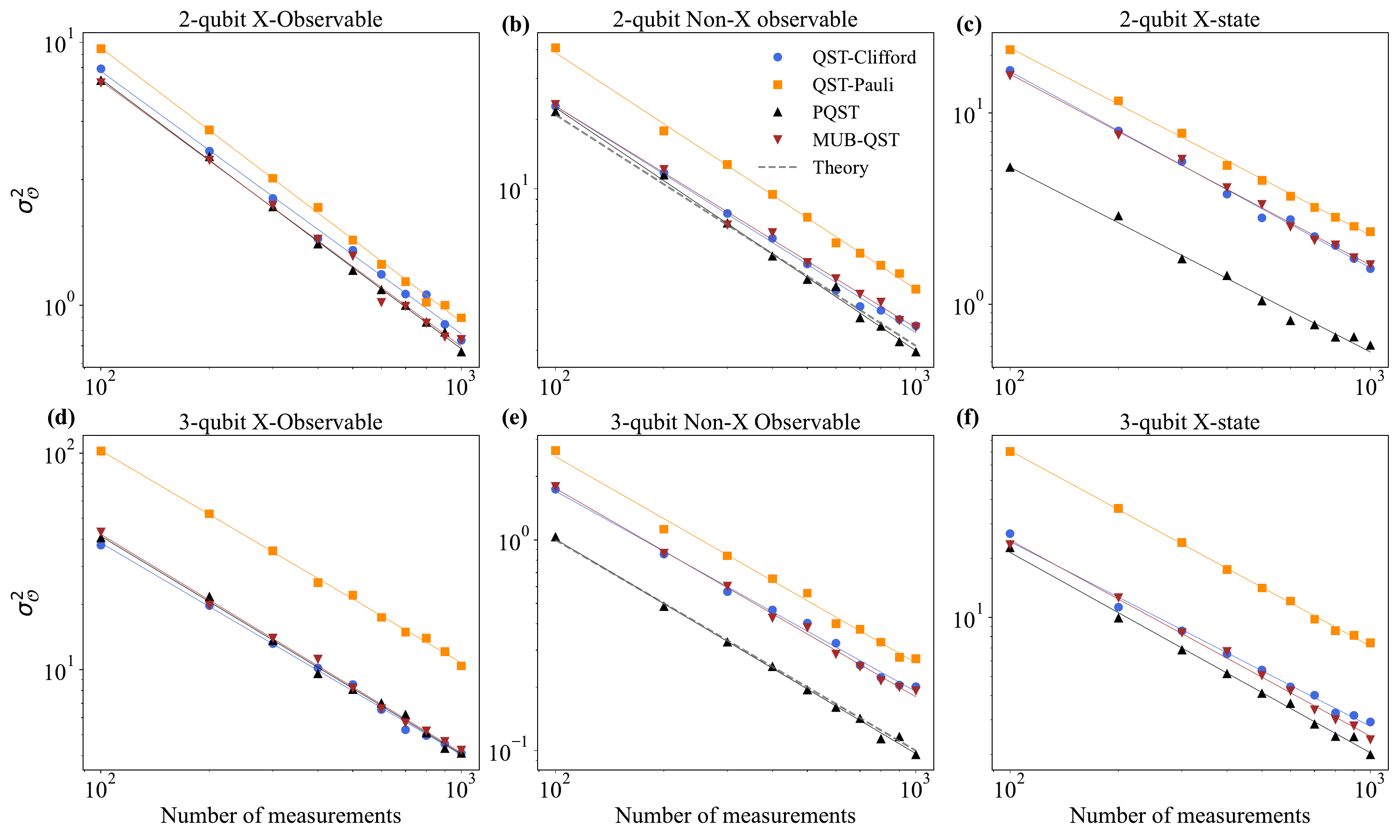}
    \caption{The scaling of MSE, $\sigma_\mathcal{O}^2$, with the number of measurements (log scale). We consider an X-type structured observable with a randomly generated quantum state for a 2-qubit system in case (a) and for a 3-qubit system in case (d). Similarly, a non-X-type observable with a randomly generated quantum state is examined for a 2-qubit system in case (b) and for a 3-qubit system in case (e). Additionally, we study the scaling behavior for a 2-qubit X-state with an arbitrarily chosen Pauli string observable in case (c) and extend this analysis to a 3-qubit X-state in case (f). Theoretical error bounds, represented by dotted lines, are shown only for non-X tomography (b,e), as in other cases they are significantly larger and therefore omitted.
}
    \label{fig:Numerical_result}
\end{figure*}
\section{Numerical examples of PQST }
\label{sec:Num}
We analyze the performance of PQST as a function of the number of measurements for different structures of the observable or the state. We then compare PQST with standard QST based on unitary 2-design Clifford sampling, unitary 1-design Pauli sampling, and measurements utilizing mutually unbiased bases (MUBs) \cite{Wang2024}. 
We evaluate the Mean Squared Error (MSE) \(\sigma_\mathcal{O}^2\) of the expectation values estimated using the PSE $\widehat{\rho}$ (generated by different methods) relative to the true expectation value, given by
\begin{align}
\sigma_{\mathcal{O}}^2 = \frac{1}{N}\sum_{i=1}^N(\Tr(\mathcal{O} \widehat{\rho}_i)-\Tr(\mathcal{O}{\rho}))^2.
\end{align} 
The results are shown in Fig.~\ref{fig:Numerical_result} (a-f). PQST achieves an equal/ improved scaling of variance with the number of measurements, compared to the standard QST using Clifford sampling, Pauli basis sampling, and MUB sampling of unitaries, thus offering lower error bounds. 
This scaling behavior has been observed for arbitrary choices of density matrices and Pauli string observables that belong to respective classes; however, we have considered three cases of randomly generated density matrices separately for a 2-qubit system and a 3-qubit system.
In certain cases, PQST even outperforms the QST using Clifford, MUB sampling, based on the specific structure of the density matrices as seen for the case of X-structured density matrices in Fig~\ref{fig:Numerical_result} (c,f). The error bounds obtained in Sec.~\ref{Sec:Sample} is consistent with numerical simulations: the actual errors are always smaller. In X-tomography, PQST offers a clear advantage over Pauli-based measurements, achieving the same scaling as Clifford and MUB sampling. For non-X tomography, PQST outperforms all other unitary sampling methods, owing to its focus on estimating elements on the $k$-active subspace.

For our numerical simulations, we have generated 2-qubit, and 3-qubit random states for a given X-structured observable (Fig. \ref{fig:Numerical_result} (a,d)) of the form in Eq.~\eqref{eq:Xstructure} as well as for non-X structured observable containing only single-active terms for 2-qubit case (Fig. \ref{fig:Numerical_result} (b)), and non-X structured observable containing only double-active terms for 3-qubit (Fig. \ref{fig:Numerical_result} (e)) case. For X-states we generate density matrices having X-structure for 2-qubit (Fig. \ref{fig:Numerical_result} (c)) and 3-qubit (Fig. \ref{fig:Numerical_result} (f)) systems, where we take the observable to be an arbitrary Pauli string operator. The exact operator values are given in Appendix~\ref{Num_detail}.
The variance is computed over $N=1000$ independent trials for each sampling size. 

\begin{figure*}
\includegraphics[width=17cm, clip=true, trim={3.5cm 9.2cm 5.8cm 4.3cm}]{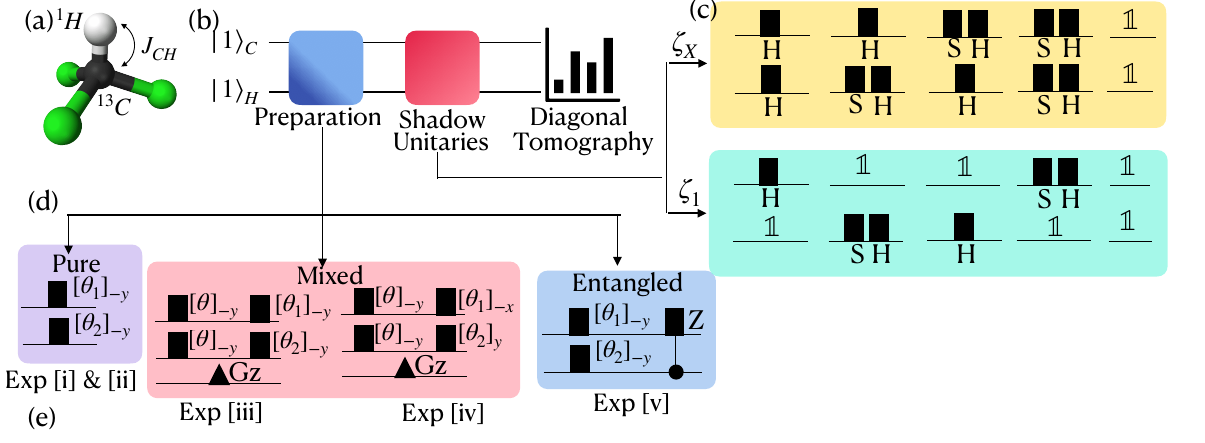}
\includegraphics[width=17cm, clip=true, trim={3.7cm 5cm 3.5cm 4.25cm}]{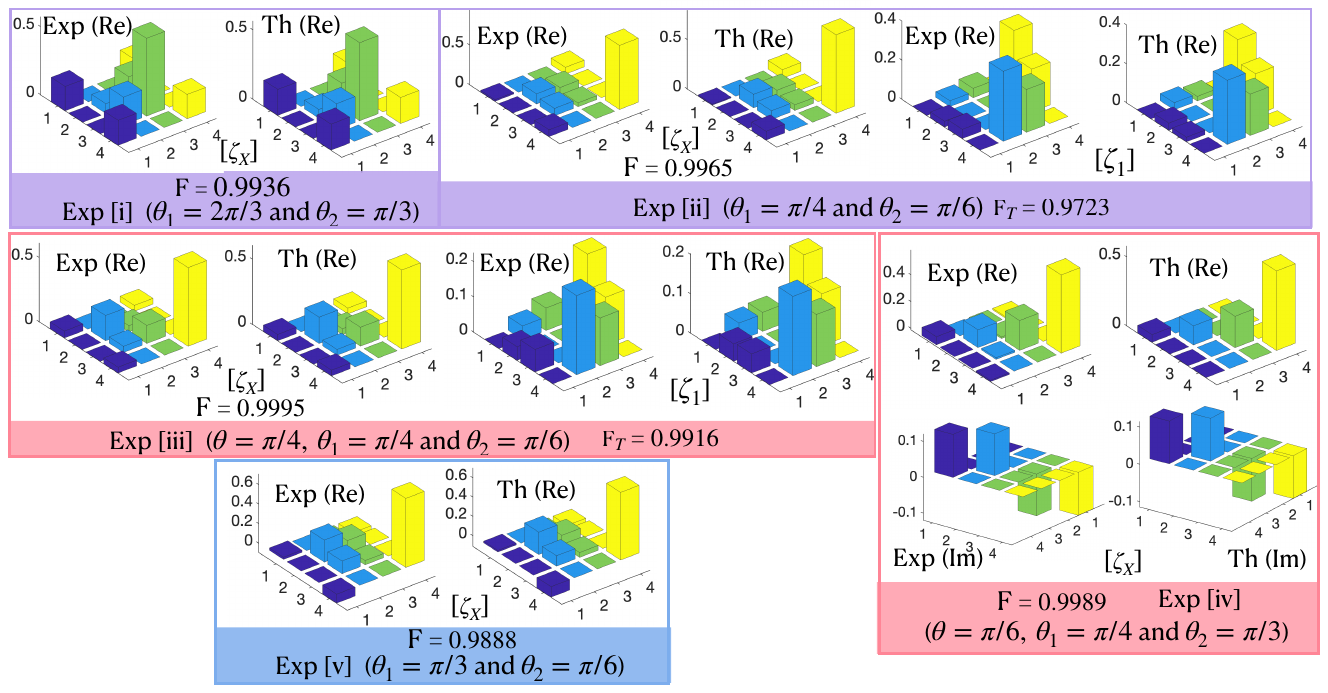}
\caption{(a) The molecular structure of 13C-Chloroform with the qubits labeled. The measured relaxation times are $T_1 = 4.88$ s, $T_2=3.5$ s and $T_2^* = 0.68$ s for $^1H$, and $T_1 = 5.78$ s and $T_2^* = 0.26$ s for $^{13}$C.
(b) A schematic showing the three basic steps involved in partial shadow tomography experiments. First, the desired state is prepared following the pulse sequences. The notation $[\theta]_p$ denotes a unitary rotation of a qubit by an angle $\theta$ about the $p$-axis ($p \in \{\pm x, \pm y, \pm z\}$) on the Bloch sphere. 
$G_z$ refers to a pulsed field gradient applied along the $z$-axis, which causes the off-diagonal elements of the density matrix to dephase, thereby reducing the purity of the quantum state. $Z$ denotes the controlled-Z gate (for more information regarding the prepared states, see Tab. \ref{tab:combined}) shown in (d), which is followed by the application of the shadow unitaries from set \(\zeta_X\) or \(\zeta_1\) (c) depending on whether we want to do an X tomography or non-X tomography. Finally, the populations are measured in the computational basis using standard diagonal tomography. The results (e) show remarkably good fidelities achieved, considering some experimental error in the preparation and applications of shadow unitaries (\(F_T\) denotes the full state fidelity, \(F\) is the fidelity of the X-shadow).} 
\label{fig:exp}
\end{figure*}
\section{Experimental Demonstration with NMR} \label{sec:exp}
\begin{table*}[]
\renewcommand{\arraystretch}{1.5}
\setlength{\tabcolsep}{5pt}
\begin{tabular}{|c|c|c|c|c|}
\hline
\multirow{2}{*}{Exp No.} & \multirow{2}{*}{State Prepared} & \multirow{2}{*}{Class} & PPS & PPS\\
 &  &  & purity & Entanglement\\
\hline
\multirow{2}{*}{Exp[i]} &  $\rho_{i}=\Ket{\eta_{\text{i}}}\Bra{\eta_{\text{i}}}, $
& \multirow{2}{*}{Pure} & \multirow{2}{*}{1} & \multirow{2}{*}{0}  \\ 
& $\Ket{\eta_{\text{i}}} = \left[\cos\left(\frac{\pi}{6}\right)\Ket{1} + \sin\left(\frac{\pi}{6}\right) \Ket{0}   \right] \otimes \left[\cos\left(\frac{\pi}{3}\right)\Ket{1} + \sin\left(\frac{\pi}{3}\right)\Ket{0}  \right]$ & & & \\
\hline
\multirow{2}{*}{Exp[ii]} &  $\rho_{ii}=\Ket{\eta_{\text{ii}}}\Bra{\eta_{\text{ii}}}, $
& \multirow{2}{*}{Pure} & \multirow{2}{*}{1} & \multirow{2}{*}{0}  \\ 
& $\Ket{\eta_{\text{ii}}} = \left[\cos\left(\frac{\pi}{8}\right)\Ket{1} + \sin\left(\frac{\pi}{8}\right) \Ket{0}   \right] \otimes \left[\cos\left(\frac{\pi}{12}\right)\Ket{1} + \sin\left(\frac{\pi}{12}\right)\Ket{0}  \right]$ & & & \\
\hline
\multirow{2}{*}{Exp[iii]} & $\rho_{\text{iii}}=\left[\frac{\mathbb{1}}{2}-\cos\left(\frac{\pi}{4} \right)R_1\right] \otimes \left[\frac{\mathbb{1}}{2}-\cos\left(\frac{\pi}{4} \right)R_2\right],  $
& \multirow{2}{*}{Mixed} & \multirow{2}{*}{0.56} & \multirow{2}{*}{0}  \\  
 & $R_1= \cos\left(\frac{\pi}{4}\right)I_z - \sin\left(\frac{\pi}{4}\right)I_x, ~~ R_2 = \cos\left(\frac{\pi}{6}\right)I_z - \sin\left(\frac{\pi}{6}\right)I_x$
&  &  &   \\  
\hline
\multirow{2}{*}{Exp[iv]} & $\rho_{\text{iv}}=\left[\frac{\mathbb{1}}{2}-\cos\left(\frac{\pi}{6} \right)R_3\right] \otimes \left[\frac{\mathbb{1}}{2}-\cos\left(\frac{\pi}{6} \right)R_4\right],$
& \multirow{2}{*}{Mixed} & \multirow{2}{*}{0.765} & \multirow{2}{*}{0}\\ 
 & $R_3 = \cos\left(\frac{\pi}{4}\right)I_z + \sin\left(\frac{\pi}{4}\right)I_y, ~~ R_4 = \cos\left(\frac{\pi}{3}\right)I_z + \sin\left(\frac{\pi}{3}\right)I_x$
&  &  & \\ 
\hline
\multirow{3}{*}{Exp[v]} & $\rho_{\text{v}}=\Ket{\eta_{\text{v}}}\Bra{\eta_{\text{v}}}, $ & \multirow{3}{*}{Entangled} & \multirow{3}{*}{1} & \multirow{3}{*}{0.28}\\
 & $\Ket{\eta_{\text{v}}}=\sin\left(\frac{\pi}{6}\right)\sin\left(\frac{\pi}{12}\right)\Ket{00} + \sin\left(\frac{\pi}{6}\right)\cos\left(\frac{\pi}{12}\right)\Ket{01} +  $ &  &  & \\
 & $~~~~~~~ \sin\left(\frac{\pi}{12}\right)\cos\left(\frac{\pi}{6}\right)\Ket{10} - \cos\left(\frac{\pi}{6}\right)\cos\left(\frac{\pi}{12}\right)\Ket{11}  $ &  &  & \\
 \hline
\end{tabular}

\caption{List of all states prepared experimentally for testing partial shadow tomography, including their purity and entanglement values. Here, purity of a density operator $\rho$ refers to $\mathrm{Tr} \, \rho^2$ (which is $1$ for pure states and $0.25$ for maximally mixed states), and entanglement is measured by usual entanglement entropy (for pure states) and by logarithmic negativity (for mixed states). Both of these measures take value $1$ for maximally entangled states and $0$ for separable states. The spin operator $I_k := \sigma_k/2$, where $\sigma_k$ is the $k$th component of the Pauli operator $\vec{\sigma}$.}
\label{tab:combined}
\end{table*}
We now describe the experimental demonstration of PQST in a two-qubit NMR register $^{13}$C-Chloroform (CHCl$_3$) wherein $^{13}$C and $^{1}$H spin-1/2 nuclei are qubit $1$ and $2$ respectively (see Fig.~\ref{fig:exp}~(a)). In a strong $\hat{z}$-magnetic field of $11.7$ T  inside a Bruker $500$ MHz NMR spectrometer, the liquid ensemble of CHCl$_3$, dissolved in Dimethyl sulfoxide (DMSO), rests in thermal equilibrium at an ambient temperature of $300$ K. Under high temperature-high field assumption  \cite{cavanagh1996protein}, the density matrix of the quantum register reads $\rho_{\mathrm{th}} = \mathbb{1}/4 + \epsilon (\gamma_{\mathrm{C}} I_z^{\mathrm{C}} + \gamma_{\mathrm{H}} I_z^{\mathrm{H}})$, where $\gamma_i$ is the gyro-magnetic ratio of the $i$'th nucleus, $I_{z}^{v}:= \hat{\sigma}^{v}_z /2$ are the spin operators, and  $\epsilon \sim 10^{-5}$ is the purity factor.  Using secular approximation in a doubly-rotating frame, rotating at the resonant frequency of each nucleus, the Hamiltonian can be written as \cite{cavanagh1996protein,levitt2008spin}
\begin{align}
\mathcal{H}_{\mathrm{NMR}} = 2\pi  J_{\mathrm{CH}} \hbar I_z^{\mathrm{C}} I_z^{\mathrm{H}},
\end{align}
where $J_\mathrm{CH}=220$ Hz is the scalar coupling constant. Starting from the thermal state $\rho_{\mathrm{th}}$, we initialize the quantum register into the pseudopure state (PPS) of $\Ket{11}$ \cite{cory1997ensemble,gershenfeld1997bulk,cory1998nuclear}. Subsequently, using the pulses shown in Fig. \ref{fig:exp}~(d), we prepare each of the five different states listed in Tab. \ref{tab:combined}.  
We now apply the shadow unitaries from the appropriate sets ($\zeta_X$ and $\zeta_1$) as shown in Fig.~\ref{fig:exp}~(c) and measure the populations via diagonal tomography.
Being an ensemble architecture, the 
NMR diagonal tomography is efficient since it only requires a single readout of the NMR signal spanning over all the spin transitions after twirling non-diagonal elements and applying a detection pulse \cite{DiagonalTomo,LEE2002349, AAST}.\\
The experimentally measured diagonal states are inverse-rotated by the same shadow unitary chosen before, then subjected to the pseudo-inverse map Eq.~\eqref{eqn:Inverse} with strength \( p = 5 \), and finally averaged over the unitary choices in each set, as described in Sec.~\ref{sec:pqstensemble}.  
The full estimator \(\widehat{\rho}\) is constructed via combining the PSEs $\widehat{\rho_i}$ generated by respective unitary sets \(\zeta_i\). 
\begin{equation}
\widehat{\rho}=\mathcal{P}_X(\widehat{\rho}_X)+\mathcal{P}_1(\widehat{\rho}_1),
\end{equation}
where \(\mathcal{P}_X(\cdot)\) preserves the elements in diagonal and anti-diagonal positions and \(\mathcal{P}_1(\cdot)\) preserves the elements in other off-diagonal (single-active) positions.
The final reconstructed states \(\widehat{\rho}\) are displayed in Fig~\ref{fig:exp}~(e) show excellent agreement with the actual states $\rho$, with most of the fidelities \(F = \left(\tr\sqrt{\sqrt{\rho}\widehat{\rho}\sqrt{\rho}}\right)^2\)
being around 0.99.  
Such high fidelities confirm the robustness of PQST against the experimental limitations in preparing the states and applying shadow unitaries in the presence of thermal electronic noise, introducing random errors and RF inhomogeneity, introducing systematic errors.

\section{Conclusion}
\label{sec:conclude}
Quantum shadow tomography is a powerful tool to estimate the expectation values of both linear and nonlinear observables for an unknown quantum state. 
However, in many settings such as in variational quantum algorithms (VQAs)~\cite{Cerezo2021}, we are interested in calculating specific expectation values such as that of nearest neighbor $X_1X_2$, $Y_1Y_2$, $Z_1Z_2$ operators in the case of a VQA to optimize the $XXZ$ ground state. Executing the full shadow tomography protocol may be unnecessary.
In such scenarios, the PQST protocol can provide an efficient alternative. 
%

We have generalized the inverse map description using the pseudo-inverse map \(\mathcal{M}^{-1}(\cdot)\) and provided a systematic approach for performing partial shadow tomography by carefully selecting subsets of single-qubit unitaries. This protocol helps to identify minimal sets of unitaries that, when paired with appropriate pseudo-inverse maps, facilitate efficient partial shadow tomography. Another convenient choice for the inverse map is an inverse depolarization map given by 
\(D_p^{-1}(A)=pA-\frac{p-1}{2^n}\mathrm{Tr}(A)\mathbb{1}\), where \(p=|\mathcal{U}|\).
This map is trace-preserving (for $\mathrm{Tr}(A) = 1$) but not completely positive. 

A promising direction for future research is the exhaustive exploration of optimal combinations of subsets of tomographically complete sets of unitaries that enable partial estimation of density matrix elements. This would allow for the estimation of subsystem properties, effectively reducing the complexity of unitary sampling while enabling more efficient subsystem shadow tomography.
PQST achieves the same power law scaling of variance \( \sigma_\mathcal{O}^2 \) with the number of measurements \( x \), \(\sigma_\mathcal{O}^2 \sim a x^{-\gamma}\) where the exponent \( \gamma \) is comparable to that obtained using Clifford or mutually unbiased basis (MUB) unitary sampling. However, for specific operators, PQST achieves a smaller amplitude $a$ of the power law, leading to a lower overall variance for the same number of measurements, as demonstrated by our numerical analysis. 
Additionally, PQST has the advantage of relying on simple local unitaries, unlike other protocols that may require nonlocal unitaries with significantly greater circuit depth. Single-qubit unitaries are easier to implement in practice, achieve better fidelities, and are noise-resistant. On account of the simplicity of the unitaries involved, our experimental implementation of PQST for the case of two NMR qubits demonstrated remarkably high fidelities.

Shadow tomography offers a robust framework for estimating expectation values, characterizing quantum states, and validating state preparation protocols. In contrast to full shadow tomography, which quickly becomes intractable, our approach exploits partial prior knowledge of the state or observables to achieve a reduced sampling complexity. These results demonstrate the advantage of shadow-based methods in structured scenarios and open the way for further refinements of state-estimation techniques that leverage partial prior knowledge.
\section{Acknowledgments}
Authors gratefully acknowledge discussions with Sai Vinjanampathy and Sooryamsh Asthana. AS and AC acknowledges Deepika Bhargava for her help in computation. GJS and TSM acknowledge funding from I-HUB QTF.

\appendix
\begin{widetext}
\section{Channel description for the 2-qubit system}
\label{2not}
The forward channel \(\mathcal{E}_d\) map generated by unitaries \(\{U\}\) sampled uniformly from the unitary set \(\zeta_d\) is given by:
\begin{align}
\mathcal{E}_d=\mathbbm{E}_{U\in\zeta_d,\hat{k}}U^\dagger\proj{\hat{k}}U\xrightarrow[\mathrm{Measurements}]{\mathrm{Large}}\mathbbm{E}_U\left[\sum_k\Bra{k}U\rho U^\dagger\ket{k}U^\dagger\proj{{k}} U\right],
\end{align}
where
\(\ket{\hat{k}}\) are the post-measurement collapsed states, \(\ket{k}\) are the computational basis and \(\mathbbm{E}[.]\) is the empirical average. The empirical average over a unitary set \(\zeta\) is computed as \(\frac{1}{|\zeta|} \sum_{U_i \in \zeta} f(U_i),\) 
where \(|\zeta|\) denotes the cardinality of the set. In the following, we have stated the results for 2-qubit system in the limit of large measurements.
We will consider the sets \(\zeta_X\) and \(\zeta_1\) as described in Sec. \ref{sec:2qpqst}. The pseudo-inverse map action Eq.~\eqref{eqn:Inverse} with \(p=5\) on each of the forward channel \(\mathcal{E}_X\) and \(\mathcal{E}_1\) generated by uniform sampling of unitaries from set \(\zeta_X\) and \(\zeta_1\), respectively is given below.
\begin{equation}
\widehat{\rho}_X=\mathcal{M}_{5}^{-1}(\mathcal{E}_X)=
\renewcommand{\arraystretch}{1.5}
\begin{pmatrix}
    \rho_{00,00} & \rho_{00,01}+\rho_{10,11} & \rho_{00,10}+\rho_{01,11} & \rho_{11,11} \\
    \rho_{01,00}+\rho_{11,10} & \rho_{01,01} & \rho_{01,10} & \rho_{01,11}+\rho_{00,10} \\
    \rho_{10,00}+\rho_{11,01} & \rho_{10,01} & \rho_{10,10} & \rho_{10,11}+\rho_{00,01} \\
    \rho_{00,11} & \rho_{11,01}+\rho_{10,00} & \rho_{11,10} + \rho_{01,00} & \rho_{11,11}  
\end{pmatrix}
\label{eqn:2qubitX}
\end{equation}
\begin{equation}
\widehat{\rho}_1=\mathcal{M}_{5}^{-1}(\mathcal{E}_1)=
\renewcommand{\arraystretch}{1.5}
\begin{pmatrix}
    2\rho_{00,00}-\rho_{11,11} & \rho_{00,01} & \rho_{00,10} & 0 \\
    \rho_{01,00} & 2\rho_{01,01}-\rho_{10,10} & 0 & \rho_{01,11} \\
    \rho_{10,00} & 0 & 2\rho_{10,10}-\rho_{01,01} & \rho_{10,11} \\
    0 & \rho_{11,01} & \rho_{11,10}  & 2\rho_{11,11}-\rho_{00,00}  
\end{pmatrix}
\label{eqn:2qubit1}
\end{equation}
The equations \eqref{eqn:2qubitX} and \eqref{eqn:2qubit1} show the extraction of subsets of elements of the full density matrix. However, if we construct PSEs using subsets of \(\zeta_{1a}\) and \(\zeta_{1b}\) using the pseudo-inverse in Eq.~\eqref{eqn:Inverse} with $p=3$ for this case, we get the following matrices:
\begin{equation}
\resizebox{\textwidth}{!}{$
\widehat{\rho}_{1a}=\mathcal{M}^{-1}_{3}(\mathcal{E}_{1a})=
\renewcommand{\arraystretch}{1.5}
\begin{pmatrix}
    -1+2\rho_{00,00}+ \rho_{10,10}& 0 & \rho_{00,10} & 0 \\
    0 & -1+2\rho_{01,01}+\rho_{11,11} & 0 & \rho_{01,11} \\
    \rho_{10,00} & 0 & -1+2\rho_{10,10}+\rho_{00,00} \\
    0 & \rho_{11,01} & 0 & -1+2\rho_{11,11}+\rho_{01,01}  
\end{pmatrix}
$}
\label{eqn:2qubit1a}
\end{equation}
\begin{equation}
\resizebox{\textwidth}{!}{$
\widehat{\rho}_{1b}=\mathcal{M}^{-1}_{3}(\mathcal{E}_{1b})=
\renewcommand{\arraystretch}{1.5}
\begin{pmatrix}
    -1+2\rho_{00,00}+\rho_{10,10}& \rho_{00,01} & 0 & 0 \\
    \rho_{01,00} & -1+2\rho_{01,01}+\rho_{00,00} & 0  & 0\\
    0 & 0 & -1+2\rho_{10,10}+\rho_{11,11} & \rho_{10,11} \\
    0 & 0 & \rho_{11,10} & -1+2 \rho_{11,11}+\rho_{10,10}
\end{pmatrix}
$}
\label{eqn:2qubit1b}
\end{equation}
The shadow estimator \(\widehat{\rho}_{H\otimes H} =-\mathbb{1}+\frac{p}{4}\mathcal{B}_H\), where \(\mathcal{B}_H\) is given by
\begin{align}
\mathcal{B}_H=\begin{pmatrix}
\begin{array}{c}
\rho_{00,00} + \rho_{01,01} \\ 
~~~~~+ \rho_{10,10} + \rho_{11,11}
\end{array} & 
\begin{array}{c}
\rho_{00,01} + \rho_{01,00} \\ 
~~~~~+ \rho_{10,11} + \rho_{11,10}
\end{array} & 
\begin{array}{c}
\rho_{00,10} + \rho_{01,11} \\ 
~~~~~+ \rho_{10,00} + \rho_{11,01}
\end{array} & 
\begin{array}{c}
\rho_{00,11} + \rho_{01,10} \\ 
~~~~~+ \rho_{10,01} + \rho_{11,00}
\end{array} \\[10pt]
\begin{array}{c}
\rho_{00,01} + \rho_{01,00} \\ 
~~~~~+ \rho_{10,11} + \rho_{11,10}
\end{array} & 
\begin{array}{c}
\rho_{00,00} + \rho_{01,01} \\ 
~~~~~+ \rho_{10,10} + \rho_{11,11}
\end{array} & 
\begin{array}{c}
\rho_{00,11} + \rho_{01,10} \\ 
~~~~~+ \rho_{10,01} + \rho_{11,00}
\end{array} & 
\begin{array}{c}
\rho_{00,10} + \rho_{01,11} \\ 
~~~~~+ \rho_{10,00} + \rho_{11,01}
\end{array} \\[10pt]
\begin{array}{c}
\rho_{00,10} + \rho_{01,11} \\ 
~~~~~+ \rho_{10,00} + \rho_{11,01}
\end{array} & 
\begin{array}{c}
\rho_{00,11} + \rho_{01,10} \\ 
~~~~~+ \rho_{10,01} + \rho_{11,00}
\end{array} & 
\begin{array}{c}
\rho_{00,00} + \rho_{01,01} \\ 
~~~~~+ \rho_{10,10} + \rho_{11,11}
\end{array} & 
\begin{array}{c}
\rho_{00,01} + \rho_{01,00} \\ 
~~~~~+ \rho_{10,11} + \rho_{11,10}
\end{array} \\[10pt]
\begin{array}{c}
\rho_{00,11} + \rho_{01,10} \\ 
~~~~~+ \rho_{10,01} + \rho_{11,00}
\end{array} & 
\begin{array}{c}
\rho_{00,10} + \rho_{01,11} \\ 
~~~~~+ \rho_{10,00} + \rho_{11,01}
\end{array} & 
\begin{array}{c}
\rho_{00,01} + \rho_{01,00} \\ 
~~~~~+ \rho_{10,11} + \rho_{11,10}
\end{array} & 
\begin{array}{c}
\rho_{00,00} + \rho_{01,01} \\ 
~~~~~+ \rho_{10,10} + \rho_{11,11}
\end{array}
\end{pmatrix}
\label{eq:HXH}
\end{align}
The shadow estimator \(\widehat{\rho}_{HS\otimes HS} =-\mathbb{1}+\frac{p}{4}\mathcal{B}_{HS}\), where \(\mathcal{B}_{HS}\) is given by
\begin{align}
 \mathcal{B}_{HS}=   \begin{pmatrix}
\begin{array}{c}
\rho_{00,00} + \rho_{01,01} \\ 
~~~~~+ \rho_{10,10} + \rho_{11,11}
\end{array} & 
\begin{array}{c}
\rho_{00,01} - \rho_{01,00} \\ 
~~~~~+ \rho_{10,11} - \rho_{11,10}
\end{array} & 
\begin{array}{c}
\rho_{00,10} + \rho_{01,11} \\ 
~~~~~- \rho_{10,00} - \rho_{11,01}
\end{array} & 
\begin{array}{c}
\rho_{00,11} - \rho_{01,10} \\ 
~~~~~- \rho_{10,01} + \rho_{11,00}
\end{array} \\[10pt]
\begin{array}{c}
-\rho_{00,01} + \rho_{01,00} \\ 
~~~~~- \rho_{10,11} + \rho_{11,10}
\end{array} & 
\begin{array}{c}
\rho_{00,00} + \rho_{01,01} \\ 
~~~~~+ \rho_{10,10} + \rho_{11,11}
\end{array} & 
\begin{array}{c}
-\rho_{00,11} + \rho_{01,10} \\ 
~~~~~+ \rho_{10,01} - \rho_{11,00}
\end{array} & 
\begin{array}{c}
\rho_{00,10} + \rho_{01,11} \\ 
~~~~~- \rho_{10,00} - \rho_{11,01}
\end{array} \\[10pt]
\begin{array}{c}
-\rho_{00,10} - \rho_{01,11} \\ 
~~~~~+ \rho_{10,00} + \rho_{11,01}
\end{array} & 
\begin{array}{c}
-\rho_{00,11} + \rho_{01,10} \\ 
~~~~~+ \rho_{10,01} - \rho_{11,00}
\end{array} & 
\begin{array}{c}
\rho_{00,00} + \rho_{01,01} \\ 
~~~~~+ \rho_{10,10} + \rho_{11,11}
\end{array} & 
\begin{array}{c}
\rho_{00,01} - \rho_{01,00} \\ 
~~~~~+ \rho_{10,11} - \rho_{11,10}
\end{array} \\[10pt]
\begin{array}{c}
\rho_{00,11} - \rho_{01,10} \\ 
~~~~~- \rho_{10,01} + \rho_{11,00}
\end{array} & 
\begin{array}{c}
-\rho_{00,10} - \rho_{01,11} \\ 
~~~~~+ \rho_{10,00} + \rho_{11,01}
\end{array} & 
\begin{array}{c}
-\rho_{00,01} + \rho_{01,00} \\ 
~~~~~- \rho_{10,11} + \rho_{11,10}
\end{array} & 
\begin{array}{c}
\rho_{00,00} + \rho_{01,01} \\ 
~~~~~+ \rho_{10,10} + \rho_{11,11}
\end{array}
\end{pmatrix}
\label{eq:HSXHS}
\end{align}

\end{widetext}
\section{General Protocol}
\label{GP}
The general protocol for an $n$-qubit system is outlined for different choices of active order terms.
\begin{itemize}
\item[(i)] \(
\zeta_X = \left\{
\mathbb{1}^{\otimes n},
\bigotimes\limits_{i=1}^n u_i
\right\}
\),
where $u_i \in \{H,HS\}$.  This set has $2^n+1$ elements and the corresponding PSE $\widehat{\rho}_X$ is generated using the pseudo-inverse with $p=|\zeta_X|=2^n+1$. \(\widehat{\rho}_X\) estimates the X-shadow, i.e., zero-active  (diagonal) and $n$-active
(anti-diagonal) elements of the density matrix.
\item[(ii)] \(
\zeta_1 = \{\mathbb{1}^{\otimes n}, \{\eta_{\alpha}\}\}  
\)
 with $\alpha \in [1,2,\cdots,n]$, 
\(
\eta_{\alpha} = \left\{
\bigotimes\limits_{i=1}^n u_i
\right\}
\),
and $u_{i=\alpha} \in \{H,HS\}$ and $u_{i\neq \alpha} = \mathbb{1}$. This set contains \( 2n+1 \) unitaries. The PSE \( \zeta_1 \) is generated via sampling these unitaries and applying the inverse map with \( p = |\zeta_1|=2n+1 \). 
The estimator \( \widehat{\rho}_1 \) estimates all the single-active elements, while the PSE \( \widehat{\rho}_{1\alpha} \), corresponding to the subset \( \zeta_{1\alpha} = \{ \mathbb{1}^{\otimes n}, \eta_{\alpha} \} \), estimates elements corresponding to the single-active of the \( \alpha \)-th qubit, with \( p = |\zeta_{1\alpha}| = 3 \). Subsequently, we can combine different subsets, \( \tilde{\zeta}_{{1}} = \bigcup\limits_i \zeta_{1i} \), and estimate the corresponding single-active terms using \( p = |\tilde{\zeta}_1| \).
\item[(iii)] \(\zeta_d = \{\mathbb{1}^{\otimes n}, \{\eta_{\alpha_1\cdots\alpha_d}\}\}\)
for distinct \(\alpha_k \in [1,2,\cdots,n]\) \(
\eta_{\alpha_1\cdots\alpha_d} = \left\{
\bigotimes\limits_{i=1}^n u_i
\right\}
\), and
$u_{i\notin \{\alpha_1,\dots,\alpha_d\}} = \mathbb{1}$, else, 
$u_{i} \in \{H,HS\}$.
In each of these subsets \(\{\eta_{\alpha_1\cdots\alpha_d}\}\), the non-identity unitaries $\{H, HS\}$ simultaneously act at $d$-chosen sites decided by the choices of \(\alpha_k\). 
There are total $\binom{n}{d}$ subsets \(\eta_{\alpha_1\dots\alpha_d}\) each with cardinality $2^d$. The PSE \(\widehat{\rho}_d\) is constructed via sampling these unitaries and the pseudo-inverse with \(p=|\zeta_d|=\binom{n}{d}\times2^d+1\). However, each subset \(\zeta_{d\alpha}\) with identity \(\{\mathbb{1}^{\otimes n},\eta_{\alpha_1\cdots\alpha_d}\}\) for a particular choice of \(\alpha_1\cdots\alpha_d\) generates the PSE \(\widehat{\rho}_{d\alpha}\), which estimates the d-active terms corresponding to the qubits chosen by \(\alpha_k\)'s, with pseudo-inverse strength \(p=|\zeta_{d\alpha}|=2^d+1\).
Similarly, we can construct PSEs using different subsets, \(\zeta_{di}\) given by \(\tilde{\zeta}_d=\bigcup\limits_i\zeta_{di}\) and tuning the pseudo-inverse  strength to be \(p=|\tilde{\zeta}_d|\).
\end{itemize}
\begin{widetext}
\section{Expectation value of the estimator of the \texorpdfstring{$\langle O \rangle$}{<O>}}
\subsection{X observable}
\label{Sec:ExpectationCal}
In this section, we show that the expectation value shown in Eq.~\eqref{eq:expectOfestimator} 
\begin{equation}
\frac{1}{p}\sum_{U\sim\mathcal{U}}\sum_{b}\langle b \vert \left(pUOU^{\dagger}-\mathrm {Tr}(O)\mathbb{1}\right) \vert b \rangle \left\langle b|U\rho U^{\dagger}|b\right\rangle 
\end{equation}
is $\langle O \rangle$. Note that the second term inside the summations gives 
\begin{equation}
\frac{1}{p}\sum_{U\sim\mathcal{U}}\sum_{b}\langle b\vert\mathrm {Tr}(O)\mathbb{1}\vert b\rangle\left\langle b|U\rho U^{\dagger}|b\right\rangle = 2^n \mathrm {Tr}(O).\label{eq:0thterm}
\end{equation}
Given that $O$ is an X-structured operator and $\rho$ is a general density matrix, these can be expanded in Pauli strings as
\begin{equation}
O = \sum_{\vec \mu\in V_\times} a_{\vec \mu} \sigma_{\vec \mu},\;\; \rho = \sum_{\vec \nu} c_{\vec \nu} \sigma_{\vec \nu},
\end{equation}
where sum over $\vec \mu\in V_\times$ represents the sum over Pauli strings from within the space $V_\times$.
Employing this expansion, we get the first term of the expectation value as 
\begin{equation}
\frac{1}{p}\sum_{U\sim\mathcal{U}}\sum_{b}\langle b\vert pUOU^{\dagger}\vert b\rangle\left\langle b|U\rho U^{\dagger}|b\right\rangle = \sum_{\mu\in V_{\times}}\sum_{\nu}\sum_{U\sim\mathcal{U}}\sum_{b}a_{\vec{\mu}}c_{\vec{\nu}}\langle b\vert U\sigma_{\vec{\mu}}U^{\dagger}\vert b\rangle\left\langle b|U\sigma_{\vec{\nu}}U^{\dagger}|b\right\rangle. \label{eq:sum1}
\end{equation}
The summand is non zero only when both Pauli strings $U\sigma_{\vec{\mu}}U^\dagger$ and $U\sigma_{\vec{\nu}}U^\dagger$ are diagonal. To simplify the discussion of the possible combinations of Pauli strings that contribute, we present the following propositions.

\begin{enumerate}
\item There is exactly one element of $\mathcal{U}=\zeta_X$ which under conjugation brings a non identity Pauli string in $V_\times$ to diagonal Pauli string i.e. in $V_d$. 

To see this, we note that for any Pauli matrix $X,Y,Z$, there is exactly one operator from $H,HS,\mathbb{1}$ which brings it to a diagonal one (i.e. in $V_d$). 
This implies, that for any Pauili string $\sigma_{\vec\mu}\in V_{ad}$ made from $X,Y$ there is a unique unitary in $\mathcal{U}$ made from products of $H,HS$ which take the $\sigma_{\vec\mu}$ to $V_d$. If the string is made of $\mathbb{1},Z$ (at least one $Z$), then $\mathbb{1}$ is the only element of $\mathcal{U}$ which makes it diagonal. Thus there is exactly one element of $\mathcal{U}$ which brings a Pauli matrix to diagonal form under conjugation.

\item  If $\sigma_{\vec \mu}$ contains at least one $Z$ and at least one $X$ or one $Y$, then $U\sigma_{\vec{\mu}} U^\dagger\notin V_d$ for any $U\in\mathcal{U}$ and any computational basis state $b$.

If $Z$ occurs on site $i$, then any unitary in $\mathcal{U}$ that contains $H$ or $HS$  on the site $i$ will rotate $Z$ to $X$ or $Y$ under conjugation. This Pauli string is not in $V_d$. Only element of $\mathcal{U}$ that does not have $H$ or $HS$ on site $i$ is $\mathbb{1}$. Since $\sigma_{\vec \mu}$ contains $X$, $\mathbb{1}\sigma_{\vec \mu}\mathbb{1}\notin V_d$.

\item  If $\sigma_{\vec \mu}$ contains $\mathbb{1}$ at least on one of the sites and $X$ or $Y$ at least on one of the sites, then $\sum_{b} \langle b|U\sigma_{\vec {\nu}}U^\dagger|b\rangle \langle b|U\sigma_{\vec {\mu}}U^\dagger|b\rangle=0$ for any $U\in \mathcal{U}$ and any $\sigma_{\vec\nu} \in V_\times$.

Let $\sigma_{\vec \mu}$  be $X$ on one of the sites $i$ and $\mathbb{1}$ on $j$. For any $U\in \mathcal{U}$ such that $U\sigma_{\vec{\mu}} U^\dagger$ is a diagonal Pauli string, $U$ must have $H$ or $HS$ on the site $i$. This implies $U$ has $H$ or $HS$ on every site. We consider the three cases for $\sigma_{\vec {\nu}}$. 
\begin{enumerate}
\item If $\sigma_{\vec {\nu}}\in V_\times$ has $Z$ on any site then $H,HS$ rotates it to $X$ or $Y$ under conjugation. So  $U\sigma_{\vec {\nu}}U^\dagger\notin V_d$.
\item If $\sigma_{\vec {\nu}}\in V_\times$ is made of $X,~Y$ on each site, then on the site $j$, $U\sigma_{\vec{\nu}}U^\dagger$ is either $X,~Y,~Z$. Since $\sum_{b} \langle b|U\sigma_{\vec {\nu}}U^\dagger|b\rangle \langle b|U\sigma_{\vec {\mu}}U^\dagger|b\rangle$ is proportional to the sum over the $j$th site basis: $\sum_{b_j=\pm 1} \langle b_j|U_j\sigma_{\nu_j}U_j^\dagger|b_j\rangle \langle b_j|U_j\mathbb{1}U_j^\dagger|b_j\rangle=\mathrm{Tr}(\sigma_{\nu_j})=0$.
\item If $\sigma_{\vec\nu}=\mathbb{1}$, then $\sum_{b} \langle b|U\sigma_{\vec {\nu}}U^\dagger|b\rangle \langle b|U\sigma_{\vec {\mu}}U^\dagger|b\rangle=\mathrm {Tr}(\sigma_{\vec\mu})$ which is $0$ as $\sigma_{\vec{\mu}}$ contains at least one non-identity Pauli matrix.
\end{enumerate}
\end{enumerate}

From these we can conclude the following:

\paragraph*{Lemma:} If $\sigma_{\vec{\nu}}\in V_{\times}$ and $\sigma_{\vec{\mu}}\notin V_{\times}$, then $\sum_{b} \langle b|U\sigma_{\vec {\nu}}U^\dagger|b\rangle \langle b|U\sigma_{\vec {\mu}}U^\dagger|b\rangle=0$.

From proposition 2, the summation of interest is 0 if $\sigma_{\vec{\mu}}$ contains $Z$ and $X/Y$. From proposition $3$, it is zero if the $\sigma_{\vec{\mu}}$ contains $\mathbb{1}$ and $X/Y$. So the summation can be non zero only if $\sigma_{\vec{\mu}}$ is a string made of only $X,Y$ or only $\mathbb{1},Z$. In other words, $\sigma_{\vec{\mu}}\in V_\times$. 

Above lemma implies that sum over $\nu$ in Eq.~\eqref{eq:sum1} can be restricted to $V_\times$
\begin{equation}
\sum_{\mu,\nu\in V_{\times}}\sum_{U\sim\mathcal{U}}\sum_{b}a_{\vec{\mu}}c_{\vec{\nu}}\langle b\vert U\sigma_{\vec{\mu}}U^{\dagger}\vert b\rangle\left\langle b|U\sigma_{\vec{\nu}}U^{\dagger}|b\right\rangle \label{eq:sum2}
\end{equation}
Considering the case of $\nu,\mu=\mathbb{1}$ separately from the rest, we get 
\begin{multline}
\sum_{U,b}\left[\sum_{\mu,\nu\in V_{\times},\neq\mathbb{1}}a_{\vec{\mu}}c_{\vec{\nu}}\langle b\vert U\sigma_{\vec{\mu}}U^{\dagger}\vert b\rangle\left\langle b|U\sigma_{\vec{\nu}}U^{\dagger}|b\right\rangle +\sum_{\mu\in V_{\times},\neq\mathbb{1}}a_{\vec{\mu}}c_{\mathbb{1}}\langle b\vert U\sigma_{\vec{\mu}}U^{\dagger}\vert b\rangle\left\langle b|U\mathbb{1}U^{\dagger}|b\right\rangle \right.\\+\left.\sum_{\nu\in V_{\times},\neq\mathbb{1}}a_{\mathbb{1}}c_{\vec{\nu}}\langle b\vert U\mathbb{1}U^{\dagger}\vert b\rangle\left\langle b|U\sigma_{\vec{\nu}}U^{\dagger}|b\right\rangle  +  \sum_{\nu\in V_{\times}}a_{\mathbb{1}}c_{\mathbb{1}}\langle b\vert U\mathbb{1}U^{\dagger}\vert b\rangle\left\langle b|U\mathbb{1}U^{\dagger}|b\right\rangle\right].
\end{multline}
This can be evaluated to be 
\begin{equation}
2^n \sum_{\mu\in V_{\times},\neq\mathbb{1}} a_{\vec{\mu}}c_{\vec{\nu}} + (2^n+1)2^na_{\mathbb{1}}c_{\mathbb{1}},
\end{equation}
where we have used proposition 1 to simplify the summation over $\nu$ and over $U$. Combining with the Eq.~\eqref{eq:0thterm}, and identifying $a_{\mathbb{1}}$ as $\mathrm {Tr}(O)/2^n$, we get the expectation value of the estimator to be 
\begin{equation}
2^n \sum_{\mu,\nu\in V_{\times}} a_{\vec{\mu}}c_{\vec{\nu}} = \mathrm {Tr}(\rho O).
\end{equation}

\subsection{\texorpdfstring{$k$}{k}-active non-X observable}
The expectation value of the estimator for $\langle O \rangle$ is given by
\label{NX-observable}
\begin{equation}
\frac{1}{p}\sum_{U\sim\mathcal{U}}\sum_{b}\langle b \vert \left(pUOU^{\dagger}-\mathrm {Tr}(O)\mathbb{1}\right) \vert b \rangle \left\langle b|U\rho U^{\dagger}|b\right\rangle, 
\end{equation}
where the unitary set corresponds to $\mathcal{U}=\zeta_k$ defined in Eq.~\eqref{eq.zetak}
Since the trace is 0 for a $k>0$-active observable, this is 
\begin{equation}
\sum_{U\sim\mathcal{U}}\sum_{b}\langle b \vert \left(UOU^{\dagger}\right) \vert b \rangle \left\langle b|U\rho U^{\dagger}|b\right\rangle = 
\sum_{\vec{\mu}\in V_{k}}\sum_{\vec{\nu}}\sum_{U\sim\mathcal{U}}\sum_{b}a_{\vec{\mu}}c_{\vec{\nu}}\langle b\vert U\sigma_{\vec{\mu}}U^{\dagger}\vert b\rangle\left\langle b|U\sigma_{\vec{\nu}}U^{\dagger}|b\right\rangle.
\end{equation}
Propositions analogous to those presented above apply, which holds in particular for \(\mathcal{U}=\zeta_k\) in the Pauli \(k\)-active subspace \(V_k\). For each $\vec{\mu}\in V_k$, there is a unique $U_{\vec{\nu}}\in \mathcal{U}$ that makes the summand non-zero. 
\begin{equation}
\sum_{\vec{\mu}\in V_{k}}\sum_{\vec{\nu}}\sum_{b}a_{\vec{\mu}}c_{\vec{\nu}}\langle b\vert U_{\vec{\mu}}\sigma_{\vec{\mu}}U_{\vec{\mu}}^{\dagger}\vert b\rangle\left\langle b|U_{\vec{\mu}}\sigma_{\vec{\nu}}U_{\vec{\mu}}^{\dagger}|b\right\rangle.
\end{equation}
The summand is non zero if $\vec \mu$ is identity inside $A(\vec{\mu})$ but can have arbitrary Pauli matrices outside $A(\vec{\mu})$. However all such terms which is not identity outside $A(\vec{\mu})$ vanishes upon summing over $b$. As a result the expectation value simplifies to 
\begin{equation}
\sum_{\vec{\nu},\vec{\mu}\in V_{k}}\sum_{b}a_{\vec{\mu}}c_{\vec{\nu}}\langle b\vert U_{\vec{\mu}}\sigma_{\vec{\mu}}U_{\vec{\mu}}^{\dagger}\vert b\rangle\left\langle b|U_{\vec{\mu}}\sigma_{\vec{\nu}}U_{\vec{\mu}}^{\dagger}|b\right\rangle +\sum_{\vec{\mu}\in V_{k}}\sum_{b}a_{\vec{\mu}}c_{\mathbb{1}}\langle b\vert U_{\vec{\mu}}\sigma_{\vec{\mu}}U_{\vec{\mu}}^{\dagger}\vert b\rangle\left\langle b|U_{\vec{\mu}}\sigma_{\mathbb{1}}U_{\vec{\mu}}^{\dagger}|b\right\rangle.
\end{equation}
Each $U\in\mathcal{U}$, diagonalizes a unique entry in $V_k$, resulting in the following expression, where we have used $\mathrm {Tr}(O)=0$ to simplify the second term above,
\begin{equation}
\mathbbm{E}[\hat{O}]=2^n\sum_{\vec{\mu}\in V_{k}}a_{\vec{\mu}}c_{\vec{\mu}}=\mathrm{Tr}(\rho O).
\end{equation}

\section{Shadow norm}
\label{Sec:Proof}
The shadow norm corresponding to a given observable \(O\) is
\begin{align}
\|{O}\|_{\text{shadow}} = \max_{\sigma:\text{state}} \left( \mathbbm{E}_{U\sim \mathcal{U}}\sum_{b \in \{0,1\}^n} \bra{b} U \sigma U^\dagger\ket{b}\bra{b}U \mathcal{M}^{-1}(O) U^\dagger\ket{b}^2 \right)^{1/2}.
\end{align}
We present a bound on this for the case of X-structured observable followed by non-X observable in the following subsections.
\subsection{X-observables}
For an X-observable given by \(O = \sum_{\vec \mu\in V_\times} a_{\vec \mu} \sigma_{\vec \mu},\)
the variance of its estimator is given by the shadow norm of the traceless part of $O$: \(O_0=O-\frac{\mathrm{Tr}(O)}{2^n}\mathbb{1}\) when sampling unitaries from \(\zeta_X\) with \(\mathcal{M}^{-1}(A)=(2^n+1)A-\Tr(A)\mathbb{1}\).
We first consider the term in the summand of the form:
\begin{equation}
\bra{b}U \mathcal{M}^{-1}(O_0) U^\dagger\ket{b} = (2^n+1)\bra{b}U O_0 U^\dagger\ket{b} - \mathrm {Tr}(O_0)=(2^n+1)\bra{b}U O_0 U^\dagger\ket{b} = (2^n+1)\sum_{\vec{\mu}\in V_{\times},{\vec{\mu}}\neq \mathbb{1} } a_{\vec{\mu}} \bra{b}U \sigma_{\vec{\mu}} U^\dagger\ket{b}.\nonumber
\end{equation}

The term \(I(U,\vec{\mu},b)=\bra{b}U \sigma_{\vec{\mu}} U^\dagger\ket{b}\) is $\pm 1$ if $U\sigma_{\vec{\mu}}U^\dagger$ is diagonal and $0$ otherwise. From the propositions listed in the previous sections, we see that $I(U,\vec{\mu},b)$ is $\pm 1$ only if either 
\begin{enumerate}
\item $\vec{\mu}\in V_{d}-\mathbb{1}$ and $U=\mathbb{1}$.
\item $\vec{\mu}\in V_{ad}$. In this case there is a unique $U\in\mathcal{U}$ which makes $I=\pm 1$. We will write $\vec{\mu}^U$ for the Pauli string that is diagonalized by $U$.
\end{enumerate}
In both cases, $I(U,\vec{\mu},b)$ gives the eigenvalue $\lambda^b_{\vec{\mu}}$ of $U \sigma_{\vec{\mu}} U^\dagger$ corresponding to the eigenvector $b$.
Putting these together, we get
\begin{align}
\frac{1}{(2^n+1)}\|O_0\|_{\text{shadow}}^2 = \sum_{\vec{\mu}\in V_{ad}} a_{\vec{\mu}}^2 + \max_{\rho}\sum_{b} \langle b \vert \rho \vert b \rangle \left |\sum_{\vec{\mu}\in V_d^{(\mathcal{S})} } a_{\vec {\mu}} \sigma_{\lambda{\vec\mu}^b}\right|^2.
\end{align}
Since $O_0$ is an X-structured operator, it can be written as a sum of diagonal and anti-diagonal parts $O_{d}, O_{ad}$. The first sum in the above expression is the Hilbert Schmidt norm $\frac{\|O_{ad}\|_{\mathrm HS}^2}{2^n}$ and the second term can be bound by noting that the $\sum_{\vec{\mu}\in V_d,\vec{\mu}\neq \mathbb{1} } a_{\vec {\mu}} \sigma_{\lambda{\vec\mu}^b}$ are the diagonal entries of $O_{0}$, and is thus upper bound by the largest eigenvalue which also is $\|O_d\|_\infty$ for a diagonal matrix. With this identification, we get
\begin{align}
\|O_0\|_{\text{shadow}} \leq  \sqrt{1+\frac{1}{2^n}} \left[  {\|O_{ad}\|_{\mathrm HS}^2}  + 2^n {\|O_{d}\|_{\infty}^2} \right]^{1/2}\leq \sqrt{2(2^n+1)}\|O\|_\infty.
\end{align}
Using the identity $ \|A\|_{\mathrm HS} \leq \sqrt{\dim(A)}\|A\|_\infty $ and Cauchy inequalities, we get that ${\mathrm RHS}\leq  \sqrt{2(2^n+1)}\|O_0\|_\infty$, thereby demonstrating that the bound is lesser than the corresponding bound on the Pauli measurements. Since these comparisons are made for the upper bounds, better insight of the relative merits are achieved by analyzing specific instances which is presented in the numerical examples in the main text.

\subsection{Non X-structure observable}
\label{Sec:NXcal}
For a $k$-active observable \(O_k=\sum_{\vec{\mu} \in V_k} a_{\vec \mu}\sigma_{\vec \mu}\). By construction \(O_k\) is traceless.
Hence, the shadow norm obtained by sampling unitaries from the unitary set \(\zeta_k\)
with pseudo-inverse map \(\mathcal{M}^{-1}(A)=pA-\mathrm{Tr}(A)\mathbb{1}\) where \(p=|\mathcal{U}|=|\zeta_k|\) is given below
\begin{equation}
\bra{b}U \mathcal{M}^{-1}(O_k) U^\dagger\ket{b} = p\bra{b}U O_k U^\dagger\ket{b} - \mathrm{Tr}(O_k)=p\bra{b}U O_k U^\dagger\ket{b} = p\sum_{\vec\mu\in V_k}\bra{b}U\sigma_{\vec\mu}U^\dagger\ket{b},
\end{equation}
as \(\mathrm{Tr}(O_k)=0.\) Similarly, to the X-observables the term $I(U, 
\sigma_{\vec\mu},b)=\bra{b}U\sigma_{\vec\mu}U^\dagger\ket{b}=\pm1$, for $U\sigma_{\vec\mu}U^\dagger$ is diagonal, otherwise $0$.
Using preposition 1, 2 on $k$-active Pauli subspace \(V_k\), if \(\vec\mu\in V_k\), \(\exists\) a unique \(U\in\zeta_k\) for each \(\sigma_{\vec{\mu}}\) s.t.  $I(U, 
\sigma_{\vec\mu},b)=\pm1$.
\begin{align}
\|O_k\|^2_\mathrm{shadow}&=\left(\frac{1}{p}\sum_{\vec\mu\in V_k}p^2a_{\vec\mu}^2\right)={p}\left(\sum_{\vec\mu\in V_k}a_{\vec\mu}^2\right),\\
\|O_k\|^2_\mathrm{shadow}&=\frac{p}{2^n}\|O_k\|^2_\mathrm{HS},\\
\|O_k\|_\mathrm{shadow}&=\sqrt{\frac{p}{2^n}}\|O_k\|_\mathrm{HS}.
\end{align}
By construction, non-X tomography allows simultaneous estimation of the expectation of the diagonal observables of the form \({O}_k^d\in\{V_d^{\vec\mu}: |A(\vec\mu)|=k\}\).
For a $k$-active Pauli string $\sigma_{\vec\mu}$, we define $V_d^{\vec\mu}$ as the space of diagonal observables that act with a $Z$ on at least one of the active sites of $\sigma_{\vec\mu}$. \(V_d^{\vec\mu} = \left\{ O_d \in V_d : \operatorname{supp}_Z(O_d) \cap \operatorname{supp}_{XY}(\sigma_{\vec\mu}) \neq \varnothing \right\}\). Here, $\operatorname{supp}_Z(O_d)$ is the set of qubits where $O_d$ acts $Z$, and $\operatorname{supp}_{XY}(\sigma_{\vec\mu})$ is the set of qubits where $\sigma_{\vec\mu}$ acts $X$ or $Y$. The variance observables of the form \(\tilde{O}=O_k^d+O_k^{ad}\) where \(O_k^{ad}\) is any linear combination of $k$-active Pauli strings is given by 
\begin{align}
\frac{1}{p}\|\tilde{O}\|_{\text{shadow}}^2 &= \sum_{\vec \mu\in V_k}a_{\vec{\mu}}^2+\max_\rho\sum_b\bra{b}\rho\ket{b}\left |\sum_{\vec{\mu}\in \{V_d ^{\vec\mu}\}} a_{\vec {\mu}} \sigma_{\lambda{\vec\mu}^b}\right|^2.\\
\|\tilde{O}\|_{\text{shadow}}^2 &\leq p\left(\frac{\|O_k^{ad}\|^2_\mathrm{HS}}{2^n}+\|O_k^d\|_\infty^2\right),\\
\|\tilde{O}\|_{\text{shadow}} &\leq \sqrt{2p}\|O_k\|_\infty.
\end{align}
The last inequality follows from the relation \(\|A\|_{\mathrm{HS}}\leq \sqrt{\mathrm{dim}(A)}\|A\|_\infty\) and Cauchy inequalities. Consequently, non-X tomography allows estimation of a broader class of observables with a slightly larger error bound.
\section{Experimental Data}
\label{sec:tomo}
\begin{figure*}
\includegraphics[width=17.2cm, clip=true, trim={0.7cm 3cm 1.2cm 2.6cm}]{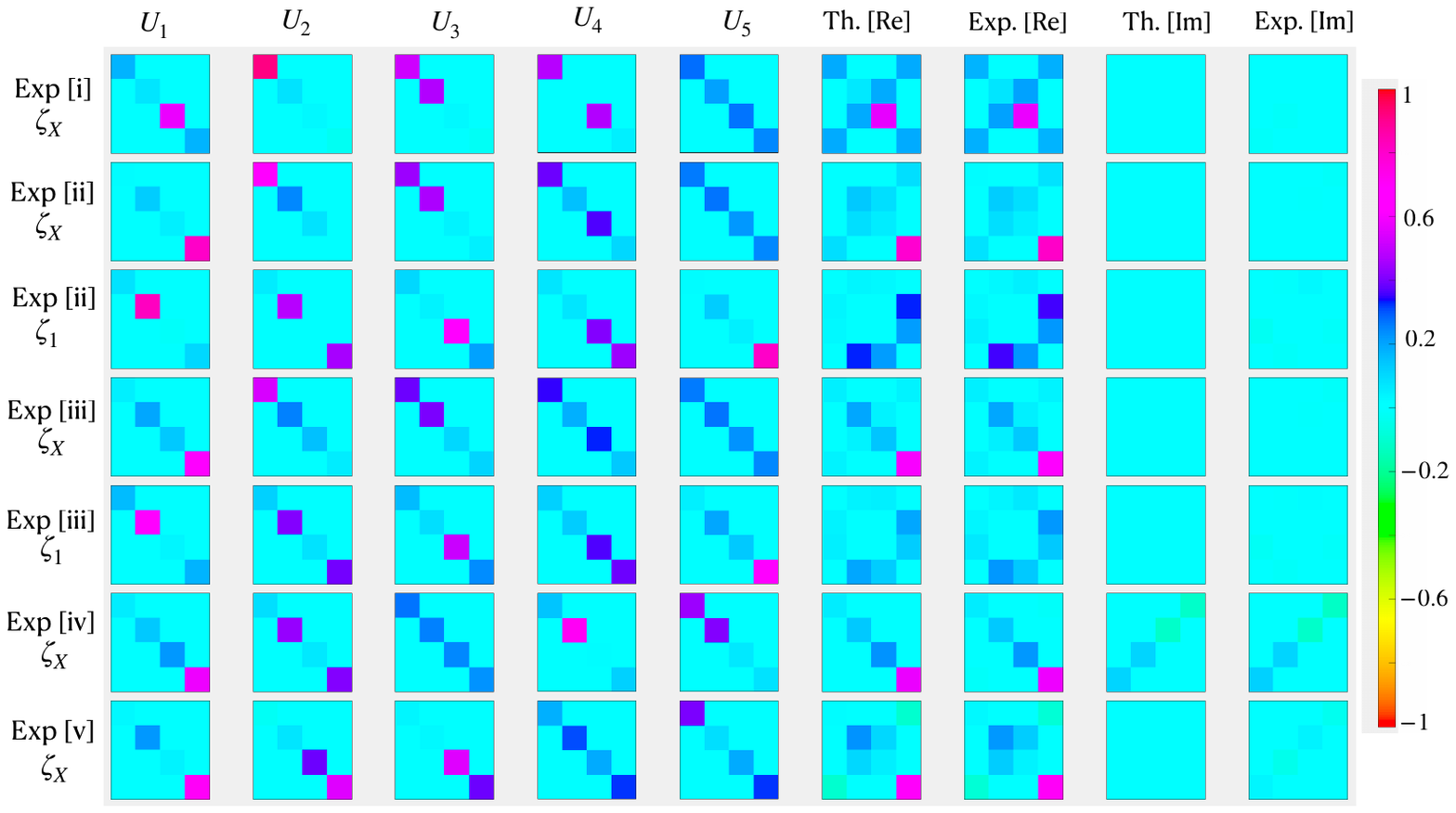}
\caption{Experimental data obtained from diagonal tomography using unitaries sampled from the sets \( \zeta_X \) and \( \zeta_1 \) for a two-qubit system. The specific unitaries \( U_i \) associated with each set are listed in Appendix~\ref{sec:tomo}. The full density matrix is reconstructed by combining partial information preserved in each estimator, \( \widehat{\rho}_X \) and \( \widehat{\rho}_1 \). The theoretical and experimental values are compared in two parts: Real: Th.[Re] and Exp.[Re] and Imaginary: Th.[Im] and Exp.[Im].
}
\label{fig:Exp_Full_Tomo}
\end{figure*}
In this section, we present the experimental data obtained from NMR diagonal tomography, as described in Section \ref{sec:exp}. Each quantum state is evolved under the action of unitaries \( U\) sampled from the sets \( \zeta_X \) and \( \zeta_1 \), followed by population measurements via diagonal tomography, as shown in Fig.~\ref{fig:Exp_Full_Tomo}. The resulting diagonal matrices are then inverse-rotated using \( U_i^\dagger \). To construct the partial state estimators (PSEs) \( \widehat{\rho}_X \) and \( \widehat{\rho}_1 \), we apply the pseudo-inverse Eq.~\eqref{eqn:Inverse} with $p=|\zeta_X|=|\zeta_1|=5$ on each of the inverse-rotated diagonally tomographed states and compute the average over all unitary choices in their respective sets. We compare the diagonal and anti-diagonal elements of the PSE $\widehat{\rho}_X$ with those of the original density matrix, and the off-diagonal elements of the PSE $\widehat{\rho}_1$ with the corresponding elements of the original density matrix by analyzing the real and imaginary parts as shown in Fig.~\ref{fig:Exp_Full_Tomo}..
\\
Finally, we reconstruct the full shadow estimator \(\widehat{\rho}\) by selectively incorporating the density matrix elements that remain preserved in each of the PSEs using Eq.~\eqref{eq:rhorecover}.
\begin{equation}
\widehat{\rho}=\mathcal{P}_X(\widehat{\rho}_X)+\mathcal{P}_1(\widehat{\rho}_1),
\end{equation}
where \(\mathcal{P}_X(\cdot)\) preserves the elements in diagonal and anti-diagonal positions and \(\mathcal{P}_1(\cdot)\) preserves the elements in other off-diagonal (single-active) positions.
The unitaries in each set are labeled in Fig.~\ref{fig:Exp_Full_Tomo} by
\begin{itemize}
    \item[] \(\zeta_X =\{U_1=H\otimes H,~U_2=H\otimes HS,~U_3=HS\otimes H,~U_4=HS\otimes HS,~U_5=\mathbb{1}\otimes\mathbb{1}\}.\)
    \item[] \(\zeta_1 = \{U_1=H\otimes\mathbb{1},~U_2=\mathbb{1}\otimes HS,~U_3=\mathbb{1}\otimes H,~U_4=HS\otimes \mathbb{1},~U_5=\mathbb{1}\otimes\mathbb{1}\}\).
\end{itemize}
\end{widetext}
\section{Numerical details}
\label{Num_detail}
In our numerical simulations in Sec.~\ref{sec:Num}, we consider randomly generated two- and three-qubit states to verify MSE (\(\sigma_\mathcal{O}^2\)) scaling with the number of measurements across different observables. Specifically, we use:  
\begin{enumerate}
    \item A two-qubit state \( \rho_2 \) to analyze MSE scaling in X-type (\( O_{2X} \)) and Non-X-type (\( O_{2NX} \)) observables.
    To compute the correct expectation values of the observables, we use \(\zeta_X\) for X-type observable and \(\zeta_1\) for non-X-type observable to generate shadow estimators \(\widehat{\rho}_X\) and \(\widehat{\rho}_1\) that accurately capture the true expectation values.\\
    \(O_{2X} = 8ZZ+2XY+3XX-10~\mathbb{1}Z\)\\
    \(O_{2NX} = 7XZ+15YZ+ 12ZX\)\\
    Note that the \( O_{2NX} \) can be converted to the X-structure via unitary transformation given by Eq.~\eqref{Transform}, with $\mathcal{U}=\mathbb{1}\otimes H$. Thus, \( O_{2NX}\rightarrow \tilde{O}_{2X}=7XX+15YX+ 12ZZ\) acquires a X-structure which can be estimated via X-shadow.

    \item A two-qubit X-state \( \rho_{2X} \) to study MSE scaling for an arbitrary observable \( O_2 \).
    We use only \(\zeta_X\) set to generate the PSE \(\widehat{\rho}_X\) to estimate the two-qubit X state.\\
    \(O_2=8ZY+12XZ+3XX-10~\mathbb{1}Z+9~\mathbb{1}\mathbb{1}\)
    
    \item A three-qubit state \( \rho_3 \) to examine MSE scaling in X-type (\( O_{3X} \)) and Non-X-type (\( O_{3NX} \)) observables. To compute the correct expectation values of the observables, we use \(\zeta_X\) for X-type observable and \(\zeta_{2c}=\zeta_{2}^{(1,3)}\) (double active on first and third qubits) for non-X-type observable to generate shadow estimators \(\widehat{\rho}_X\) and \(\widehat{\rho}_{2c}\) that accurately capture the true expectation values.\\
    \(O_{3X}=2 ~\mathbb{1} \mathbb{1} Z + 16 XXX + 6 XYX + 8 YYX + 10~\mathbb{1} ZZ \)\\ 
    \(O_{3NX}=2 XZY + 4 Y\mathbb{1}Y\) 
    
    \item A three-qubit X-state \( \rho_{3X} \) to evaluate MSE scaling for an arbitrary observable \( O_3 \).\\
    \(O_{3}=5 XXX + 10 ZZZ + 7 XYY - 6 Z \mathbb{1} Z + 6 YYY + 7 ZXX - 2 ZX \mathbb{1} \)
\end{enumerate}

This structured approach systematically compares MSE scaling behaviors across different structured states and observable types.
\begin{widetext}
\begin{equation}
\renewcommand{\arraystretch}{1.5}
    \rho_2=
\begin{pmatrix}
0.3484 & 0.0242 + 0.1014i & 0.0118 - 0.0301i & -0.1986 + 0.0933i \\
0.0242 - 0.1014i & 0.2641 & 0.0447 - 0.0050i & -0.0548 - 0.0516i \\
0.0118 + 0.0301i & 0.0447 + 0.0050i & 0.1210 & 0.0263 - 0.0367i \\
-0.1986 - 0.0933i & -0.0548 + 0.0516i & 0.0263 + 0.0367i & 0.2665
\end{pmatrix}
\end{equation}
\begin{equation}
\renewcommand{\arraystretch}{1.5}
\rho_{2X} =
\begin{pmatrix}
0.19375 & 0 & 0 & 0.09375 \\
0 & 0.30625 & -0.20625 & 0 \\
0 & -0.20625 & 0.30625 & 0 \\
0.09375 & 0 & 0 & 0.19375
\end{pmatrix}
\end{equation}

\begin{equation}
\resizebox{\textwidth}{!}{$
\renewcommand{\arraystretch}{1.5}
    \rho_3=
\begin{pmatrix}
0.1855 & -0.0429+0.0097i & 0.0075-0.0288i & 0.0319-0.0305i & -0.0640-0.0150i & 0.0061+0.0318i & -0.0125-0.0371i & 0.0348-0.0563i \\
-0.0429-0.0097i & 0.1172 & 0.0383+0.0321i & 0.0171-0.0024i & 0.0434-0.0252i & 0.0786-0.0181i & -0.0078+0.0359i & -0.0350+0.0078i \\
0.0075+0.0288i & 0.0383-0.0321i & 0.1012 & 0.0545-0.0414i & 0.0106-0.0673i & 0.0505-0.0307i & 0.0487-0.0143i & -0.0449+0.0372i \\
0.0319+0.0305i & 0.0171+0.0024i & 0.0545+0.0414i & 0.0957 & 0.0118-0.0219i & 0.0630+0.0153i & 0.0474-0.0341i & -0.0510+0.0032i \\
-0.0640+0.0150i & 0.0434+0.0252i & 0.0106+0.0673i & 0.0118+0.0219i & 0.1038 & 0.0349+0.0267i & -0.0042+0.0408i & -0.0387-0.0013i \\
0.0061-0.0318i & 0.0786+0.0181i & 0.0505+0.0307i & 0.0630-0.0153i & 0.0349-0.0267i & 0.1308 & 0.0294-0.0356i & -0.0518+0.0164i \\
-0.0125+0.0371i & -0.0078-0.0359i & 0.0487+0.0143i & 0.0474+0.0341i & -0.0042-0.0408i & 0.0294+0.0356i & 0.1359 & -0.0453+0.0288i \\
0.0348+0.0563i & -0.0350-0.0078i & -0.0449-0.0372i & -0.0510-0.0032i & -0.0387+0.0013i & -0.0518-0.0164i & -0.0453-0.0288i & 0.1300
\end{pmatrix}
$}
\end{equation}
\begin{equation}
\renewcommand{\arraystretch}{1.5}
    \rho_{3X}=
\begin{pmatrix}
0.20 & 0    & 0    & 0    & 0    & 0    & 0    & 0.05+0.02i \\
0    & 0.15 & 0    & 0    & 0    & 0    & 0.04+0.03i & 0 \\
0    & 0    & 0.10 & 0    & 0    & 0.03+0.01i & 0    & 0 \\
0    & 0    & 0    & 0.18 & 0.06+0.02i & 0    & 0    & 0 \\
0    & 0    & 0    & 0.06-0.02i & 0.12 & 0    & 0    & 0 \\
0    & 0    & 0.03-0.01i & 0    & 0    & 0.10 & 0    & 0 \\
0    & 0.04-0.03i & 0    & 0    & 0    & 0    & 0.08 & 0 \\
0.05-0.02i & 0    & 0    & 0    & 0    & 0    & 0    & 0.07
\end{pmatrix}
\end{equation}

\end{widetext}

\bibliography{bibliography}
\end{document}